\documentclass[12pt]{article}

\usepackage{epsfig}

\begin{document}

\baselineskip=20pt

\begin{Large}
\noindent \vspace{0.5cm}

\noindent{\bf Searching for a solution to the cosmological 
              constant problem -- a toy model}
\end{Large}

\bigskip	
\bigskip

\baselineskip=0.206in

\begin{large}
\noindent Wojciech Tarkowski 
\end{large}

\bigskip

\noindent {\it St.\ Paul Research Laboratory, ul.\ \.Zega\'nska 24c/2,
04-713 Warsaw, Poland
\\ e-mail:}\ {\tt tarkowski@data.pl}

\sloppy
\date{}

\bigskip
\bigskip
\bigskip

\noindent {\bf Abstract}

\medskip

\noindent {\small 
This paper concerns the so-called cosmological constant problem. In
order to solve it, we propose a toy model providing an extension of the
dimensionality of spacetime, with an additional spatial dimension
which is macroscopically unobservable. The toy model introduces no 
corrections to most predictions of the
``standard'' general relativity regarding, among others, the so-called
``five tests of general relativity''.  However, it seems that the toy
model could provide an explanation to the flatness of circular velocity
curves of spiral galaxies without introducing any dark matter.
The proposed model has quite important cosmological
consequences.  By introducing certain corrections to Friedmann's
currently accepted model(s), the toy model allows one to solve problems
related to the present density of matter in the Universe and, finally,
it does not contain the initial singularity.}

\bigskip
\bigskip
\bigskip
\bigskip
\bigskip
\bigskip
\bigskip
\bigskip
\bigskip
\bigskip
\bigskip

\baselineskip=0.18in


\smallskip

\vspace{0.5cm}

\baselineskip=0.20in

\newpage

\section{Introduction}

There has probably never existed such a difference between predictions
given by two commonly approved and powerful theories. On the one hand,
astronomical observations place strong limits on the value of the
cosmological constant $\lambda$ requiring it should not be greater than
$10^{-52}$ m$^{-2}$; see Ref.~\cite{Peebles}.  On the other hand, the
quantum theory predicts that anything contributing to the vacuum-energy
density should act like a cosmological constant. Theoretical
expectations thus give $\lambda$ of the order of $10^{70}$ m$^{-2}$, 
which exceeds observational limits by about $120$ orders of magnitude. 
This huge discrepancy is at the origin of a dilemma often referred to 
as the cosmological constant problem. Recently, many attempts have been
undertaken to solve this paradox; see Ref.~\cite{w}.  In this paper, we
propose a new and very simple solution to it, which will be called a
toy model.

\section{Einstein generalized equation}

We start this work by writing the Einstein generalized field equation,
\begin{equation}
R_{\mu\nu} - \frac{1}{2} g_{\mu\nu} R + \lambda g_{\mu\nu} =
\kappa T_{\mu\nu} \ ,
\label{1}
\end{equation}
where $R_{\mu\nu}$, $g_{\mu\nu}$ and $T_{\mu\nu}$, with $0\le
\mu,\nu\le k \in {\cal N}$, are the components of the Ricci, metric and
stress--energy tensors, respectively, $\lambda$ denotes the cosmological
constant, $R\equiv R_{\mu\nu} g^{\mu\nu}$, and $\kappa = - 8 \pi G /
c^4$; we also assume that the relationship between the Ricci and
Riemann--Christoffel tensors is of the form $R_{\mu\nu} \equiv
{R^\alpha}_{\mu\nu\alpha}$. Note that in our
approach we consider the term $\lambda g_{\mu\nu}$ in Eq.~(\ref{1}) 
to be related to the left-hand side, i.e.\ the geometric side of
the Einstein equation. Therefore, we believe that this term does not
contribute to the stress--energy tensor $T_{\mu\nu}$.

It is clear that, assuming the value of the constant $\lambda$ to be
of the order of $10^{70}$ m$^{-2}$ in a four-dimensional $(1+3)$ 
spacetime, we are not
able to obtain any reasonable, i.e.\ consistent with observations,
solution of Eq.~(\ref{1}). One of the possible ways out of this
situation is to increase the spacetime dimensionality. Yet, how many
additional dimensions are needed in such a case? We will answer this
question by defining the vacuum stress--energy tensor $T_{\mu\nu}$ which
has to render appropriate -- from the point of view of the quantum
field theory -- parameters of the vacuum, i.e.\ among others its
enormously large density of energy.  To do this, let us appeal to the
Newtonian gravity theory, which remains an excellent approach to
Einstein's general relativity at relatively small distances and
velocities,\footnote{Note, however, that the Newtonian theory remains
valid also while describing the relatively slow {\it mean} motion of a
relativistic fluid, i.e.\ the fluid whose (most) particles move with
velocities which are comparable to the velocity of light.} and note
that -- according to experimental data -- the generalized Poisson
equation \cite{Peebles} for the Newtonian gravitational potential
$\psi$ in a small region of an empty space should take the
form $\nabla_{\bf r}^2 \psi \approx 0$ where ${\bf r} \equiv
[x, y, z]$.  Since for the generalized 
Poisson equation one has $\nabla_{\bf r}^2 \psi
\propto {\rm Tr} ({\rm diag} \, T_{\mu\nu})$ so, consequently, 
${\rm Tr}({\rm diag} \, T_{\mu\nu}) \approx 0$ must occur, 
which -- due to the large value of the constant $\lambda$ --
obviously rules out simple solutions of the type of 
$T_{\mu\nu} = \pm \lambda g_{\mu\nu}/\kappa$ for all $\mu$ and $\nu$.  
At the same time we require space with the three spatial 
dimensions -- the only ones which are subject to our direct 
perception -- to be homogeneous, isotropic, and finally, to show 
a sufficiently small curvature for relatively short distances.
Moreover, the additional spatial dimension(s) should not be
directly observable. It is easy to demonstrate that the stress--energy
tensor complying with all the above requirements should, after
diagonalization, take the form $({\rm diag} \, T_{\mu\nu}) = 
(X, -X, 0, 0, 0)$ -- with an appropriately adjusted value of the 
quantity $X \in {\cal R}$ --
which can be written as $({\rm diag} \, T_{\mu\nu}) = (u_{\it
vac}, {\widetilde{p}}_{\it vac}, 0, 0, 0)$, where we define the 
quantity $u_{\it vac}$ as a
vacuum-energy density and the quantity ${\widetilde{p}}_{\it vac}$ 
-- as a vacuum pressure. Then, the only non-vanishing components 
of the stress--energy tensor $T_{\mu\nu}$ are
\begin{equation}
T_{\mu\nu} = \frac{\lambda}{\kappa} g_{\mu\nu} \qquad \hspace{0.03cm}
\mbox{for $\;$ $\mu = 0 ,1$} \;\;\; \hspace{0.04cm} \& \;\;\;
\mbox{$\nu = 0, 1.$}
\label{t} 
\end{equation}
Hence, the spacetime of our toy model is five-dimensional $(1 + 4)$,
of the form ${\cal R}^1 (\mbox{time}) \times {\cal R}^1 (\mbox{extra
spatial dimension}) \times {\cal R}^3 (\mbox{three-dimensional
space})$.  For this and subsequent sections of this paper, 
the signature of the
metric tensor is assumed to be equal to $-3$, so the elements of the
diagonalized metric tensor have the signs $(+,-,-,-,-)$.  We also
assume that the cosmological constant has a negative value, and that
$|\lambda| \sim 10^{70}$ m$^{-2}$.
Consequently, the value of the vacuum-energy density
\begin{equation}
u_{\it vac} = - \frac{c^4 \lambda}{8\pi G} \ ,
\label{g}
\end{equation}
where $G$ denotes the Newtonian gravitational constant and $c$ stands
for the speed of light, is of the same order of magnitude as that 
predicted by the quantum theory.  Note that $T_{\mu\nu}$, for 
$\mu = 0, 1$ and $\nu = 0, 1$, is the stress--energy
tensor of an ideal fluid which satisfies the vacuum equation of state,
$u_{\it vac} = - {\widetilde{p}}_{\it vac}$, and the (negative)
vacuum pressure
\begin{equation}
{\widetilde{p}}_{\it vac} = \frac{c^4 \lambda}{8\pi G}
\label{p}
\end{equation}
comes exclusively from the additional spatial dimension. In other
words, the pressure ${\widetilde{p}}_{\it vac}$ acts along the extra
spatial dimension (further denoted by $a$), i.e.\ orthogonally (on)to
hypersurfaces with constant values of the coordinate $a$.  Pressure
along the three ``macroscopic'' spatial dimensions is equal to zero.

Note also that for Eq.~(\ref{1}) with the stress--energy tensor defined
by expression (\ref{t}), the corresponding generalized Poisson equation 
takes the form $\nabla_{\bf r}^2 \psi = - \kappa c^2 (u_{\it vac} +
{\widetilde{p}}_{\it vac})/2 = 0$, according to our expectations and to
the requirement imposed while constructing the stress--energy tensor
$T_{\mu\nu}$ at the beginning of this section; see also section 5.1 
of this paper.

It is worth emphasizing that the Einstein equation (\ref{1}) with the
stress--energy tensor given by expression (\ref{t}) implies 
that the only non-vanishing components 
of the Ricci tensor $R_{\mu\nu}$ are
\begin{equation}
R_{\mu\nu} = \lambda g_{\mu\nu} \qquad \mbox{for $\;$ $\mu = 0, 1$}
\;\;\; \hspace{0.04cm} \& \;\;\; \mbox{$\nu = 0, 1.$}
\label{re}
\end{equation}

\section{Solution for an empty space}

The solution of Eq.~(\ref{1}) with the
stress--energy tensor given by expression (\ref{t}) reads
\begin{eqnarray}
ds^2 = \left(1 + |\lambda| a^2\right)\!  c^2 dt^2 - \left( 1 +
|\lambda| a^2\right)^{\! -1} da^2 - d x^2 - d y^2 - d z^2
\label{2}
\end{eqnarray}
where $a\equiv x^1$ and $(x , y , z) \equiv (x^2 , x^3 , x^4)$, with
$x^1$ and $(x^2 , x^3 , x^4)$ denoting the ``micro'' and macrospace
coordinates, respectively.\footnote{The term ``microspace'' is
used here to emphasize that the additional spatial dimension is
macroscopically, or directly unobservable, which will be shown later.}

Let us first investigate the properties of the metric (\ref{2}) on a
``microscale''. Taking $(x, y, z) = {\it const}$ we obtain
\begin{equation}
ds^2 = \left(1 + |\lambda| a^2\right)\!  c^2 dt^2 - \left( 1 +
|\lambda| a^2\right)^{\! -1} da^2 \ .
\label{3}
\end{equation}
This metric describes the covering surface ${\cal R}^1 \times {\cal
R}^1$ of the anti-de Sitter two-dimensional spacetime with the
negative cosmological constant $\lambda$. It is known in turn, that the
anti-de Sitter spacetime of the form ${\cal S}^1 \times {\cal R}^1$
has no Cauchy surfaces and contains ``global'' closed time-like curves
\cite{Hawking,Thirring}. Namely, after a (coordinate) time
\begin{equation}
T_l = \frac{2 \pi}{c \sqrt{|\lambda|}}
\label{4}
\end{equation}
has elapsed, the observer located at any place where the condition $a=
{\it const}$ is satisfied, would retrace his own life history (of
course, the covering surface given by the metric (\ref{3}) does not
possess this property any more). Hence, one obtains a ``natural'' unit
imposed on the anti-de Sitter spacetime:\ the coordinate time $T_l \sim
10^{-43}$ s. Note that the maximum distance, which can be covered
during the time $T_l$ by a signal or a particle, is equal to 
$L \equiv c \, T_l \sim 10^{-35}$ m. 
In the next section it will be shown that no particle with a finite
energy, moving in the spacetime described by expression (\ref{2}), could
irrevocably leave the nearest neighbourhood $a \sim \pm L$ of the
macrospace given by $a=0$. First, however, we should point out that
the quantity $T_l$ is of the order of the Planck time $T_{\it Pl} =
(G\hbar/c^5)^{\! 1/2}$, where $\hbar \equiv h/(2 \pi)$ denotes the Planck
constant, and $L$ is of the order of the Planck length $L_{\it Pl} =
(G\hbar/c^3)^{\! 1/2}$. Bearing in mind the formula (\ref{4}), one then
can assume that in our toy model the value of the cosmological constant
is given by the combination of the fundamental constants of nature,
\begin{equation}
\lambda = - \frac{c^3}{G \hbar} \cong - 3.829 \times 10^{69}
\;\, {\rm m^{-2}}.
\end{equation}

\subsection{Equations of motion}

Now we will present considerations which lead to the conclusion that
the additional spatial dimension is directly unobservable.  Taking the
metric (\ref{2}) we obtain integrated equations of the motion for a 
massless as well as an ordinary (i.e.\ possessing a finite mass) test
particle\footnote{Note that in this paper we investigate the behaviour
of uncharged particles only.} in the absence of any potentials,
\begin{eqnarray}
\left( \frac{d a}{d\gamma}\right)^{\! 2} &=&
U^2 - \left( 1+ |\lambda| a^2 \right) |{\bf p}|^2
\label{e1} \\
m^2 \!\left( \frac{d a}{d\sigma}\right)^{\! 2} &=& U^2 - \left( 1 +
|\lambda| a^2 \right)
\!\left(m^2 + |{\bf p}|^2 \right) \ ,
\label{e2}
\end{eqnarray}
respectively. As usual, the quantities $\gamma$ and $\sigma$ denote 
here affine parameters; in particular, the parameter $\sigma$ may be
equal to the proper time $\tau$ of a particle with a finite mass.
The constant quantities $m$, $U$ and 
${\bf p}$ are the particle's rest-mass,
total energy and three-dimensional momentum vector in the
``macroscopic'' space $a=0$, respectively, and $c$ denotes the velocity
of a massless particle with regard to the hypersurface $a=0$. 
Finally, $U \equiv p_0$, $|{\bf p}|^2 = 
p^2 \equiv - \sum_{i=2}^4 p_i p^i =\sum_{i=2}^4 (p^i)^2$, 
and $p^a \equiv p^1$ is equal to 
the square root of the left-hand side of
Eqs.~(\ref{e1}) or (\ref{e2}).  It is then easy to conclude that only
particles with huge energies $U$ are able to surmount the incredibly
deep potential well\footnote{Of course, this is true only for particles
which fulfil the condition $p \neq 0$ or $m \neq 0$.  Otherwise, 
i.e.\ if $p = 0$ and $m = 0$, one has $a \to \pm \infty$ even for an
arbitrarily small (but non-zero) value of the energy $U$; 
see Eqs.~(\ref{e1}) and/or (\ref{e2}).} -- proportional to 
$( 1+ |\lambda| a^2 )^{\! 1/2}$, and penetrate the additional 
spatial dimension $a$ on a scale much larger than 
$L \sim |\lambda|^{-1/2}$.(\footnote{See also Ref.~\cite{Rubakov}.
The authors conclude that the extra dimension(s), even though
non-compactified, would be unobservable directly, if ordinary
``three-dimensional'' (light) particles were confined inside a
potential well which is narrow enough along the additional spatial
dimension(s), but flat along the three physical ones. As we have seen,
for the spacetime with the metric (\ref{2}), the two above
requirements are satisfied.}) Moreover, there exists only a single
stable stationary point for the particle moving in
the additional spatial dimension, given by
\begin{equation}
0 = \frac{d}{da} \sqrt{ 1 + |\lambda| a^2} \ ;
\label{c}
\end{equation}
it is of course $a=0$, i.e.\ the ``ordinary'' four-dimensional
``macroscopic'' spacetime. The acceleration of any object moving
towards this point (or, to be more precise, this hyperspace) would be
enormously large even for extremely short distances $a$.

Assuming the right-hand sides of Eqs.~(\ref{e1}) and (\ref{e2}) to be
non-negative, one may solve both of them, obtaining
\begin{eqnarray}
a(\gamma) &=& \pm \sqrt{\frac{U^2 - p^2}{p^2 |\lambda|}}\,
\sin\!\left( p \sqrt{|\lambda|}\,
\gamma\right)
\label{r1} \\
a(\eta) &=& \pm \sqrt{\frac{U^2 -
\left(m^2+p^2\right)}{\left(m^2+p^2\right)
|\lambda|}}\, \sin\!\left[ \sqrt{\left(m^2+p^2\right)|\lambda|}\,
\eta \right] \ ;
\label{r2}
\end{eqnarray}
the quantity $\eta \equiv \sigma/m$ in expression (\ref{r2}) is the 
affine parameter.

First let us focus on the formula (\ref{r1}).  Comparing $p \,
[G|\lambda|/(\hbar c)]^{1/2} = \omega$ for the argument of the sine
function (the momentum $p\equiv |{\bf p}|$ is expressed here in SI
units) and substituting $\lambda = - c^3/(G\hbar)$, we arrive at the
fundamental quantum relation between the momentum and the 
wave frequency, $p=\hbar\omega/c$.  However, this formula now has a
completely new meaning:\ the ``ordinary'' three-dimensional momentum
of a massless particle is proportional to the frequency of its
oscillations in the additional spatial dimension. It should be pointed
out that this relationship results exclusively from the spacetime
geometry.

Now let us define $E$ conventionally 
as the energy of a particle in the
four-dimensional special-relativistic spacetime $a=0$ in the absence
of the additional spatial dimension; it will further be demonstrated
that in the toy model, for the case of a particle or of a system of
particles moving along the geodesic line(s), this quantity remains
quasi-constant for sufficiently long time intervals. Since $E \equiv c
p$, we arrive at the familiar expression called the Planck formula,
$E=\hbar \omega$.  Similarly, in the case of expression (\ref{r2}), we 
have $E^2 \equiv m^2 c^4 + c^2 p^2 = \hbar^2 \omega^2$. 
For the four-dimensional coordinate system in the spacetime given by
$a=0$, in which the particle rests ($p=0$), we may then write $E_0 =
mc^2 = \hbar \omega_0$, where $\omega_0$ is the oscillation (or wave) 
frequency of the particle in this system. The formula
\begin{equation}
E^2 - c^2 {\bf p}^2 = m^2 c^4 = \hbar^2 \omega^2_0 =
\hbar^2 \omega^2 - \hbar^2 c^2
{\bf k}^2 \ ,
\label{fv}
\end{equation}
where $\bf k$ denotes the three-dimensional wave-vector
in the ``macroscopic'' space $a = 0$, is satisfied in another
coordinate system in which the particle moves. Here, for a moment, we
made use of Einstein's special relativity theory which corresponds to
the spacetime determined by $a=0$ in the metric (\ref{2}).  Special
relativity explicitly and unambiguously determines the extension of the
Planck formula $E = \hbar \omega$ on four-dimensional vectors present
in this theory. Namely, in special relativity both 
the structures:\ $(E/c,
{\bf p})$ and $(\omega/c, {\bf k})$ form the four-vectors. The energy
$E$ in the four-vector is accompanied by the three momentum components 
while the wave frequency $\omega$ goes with the three components of the
wave-vector $\bf k$. The formula (\ref{fv}) reflects the fact that both
the four-vectors have constant lengths in different coordinate systems 
of the spacetime $a = 0$. The particle's 
momentum vector $\bf p$ is in the
direction of the oscillation, or wave normal, so it has
the same direction as the wave-vector $\bf k$; thus, comparing
$E = \hbar\omega$ in Eq.~(\ref{fv}) we arrive at the de~Broglie 
relation ${\bf p} = \hbar {\bf k}$.

Let us note that the trajectory of a particle is, in fact, identical 
to the trajectory of a certain (uniform) transverse plane wave
which propagates in 
the ``macroscopic'' three-dimensional space $a = 0$; see the 
formulae (\ref{r1}) and/or (\ref{r2}).  In other words, one can regard 
the particle's trajectory as a trajectory 
of a certain wave. In turn, we know
that the simplest plane wave is characterized by the wave frequency
$\omega$ and by the wave-vector $\bf k$ that has the direction
coincident with the direction of the wave propagation and the length
$2\pi/\lambda$, where $\lambda$ denotes the wavelength. Such a wave may
be described by the following representation
\begin{equation}
B \exp [ i ( \pm {\bf k}\cdot{\bf r} - \omega \, t ) ]
\label{repr}
\end{equation}
where $B$ denotes the wave amplitude, $i$ is the imaginary unit, and 
${\bf r} \equiv \left[x^2,x^3,x^4\right] = \left[x_2,x_3,x_4\right]$.
On the other hand, for the
discussed particle we have $\omega = E/\hbar$ and ${\bf k} = {\bf
p}/\hbar$. Hence, the simplest wave representation of any
(free\footnote{Note that expression (\ref{2}) is actually a generalized
special-relativity metric, since it incorporates no gravitational
effects (from masses) or other interactions.}) particle can 
be written as
\begin{equation}
B \exp \!\left[ \frac{i}{\hbar} \!\left( \pm
{\bf p}\cdot{\bf r} - E \, t \right) \right]
\equiv D \!\left( {\bf r} \right)
\exp \!\left(- \frac{i}{\hbar} E \, t \right) \ ,
\label{waver}
\end{equation}
where the function
$D({\bf r})$ denotes the spatial part of the wave representation.  Note
that the complex time factor $\exp(- i E t/\hbar)
\equiv \exp(- i \omega t)$ or the linear superposition of such factors
with different oscillation frequencies $\omega$, present in the above
representation and in most wave-functions of quantum mechanics 
(i.e.\ for the cases of time-independent external fields, or stationary
states), obtains simple interpretation in the toy model. It just
corresponds to the oscillations of the investigated object in the
additional spatial dimension. 

Let us proceed with the analysis of the relationships (\ref{r1}) and
(\ref{r2}). In particular, assuming that $E=\hbar\omega=0$, from the 
formulae (\ref{r1}) or (\ref{r2}) we obtain $a=\pm U \gamma$ or 
$a = \pm U \eta$, respectively. This means that particles satisfying 
the conditions $E=0$ (or $\omega=0$) and $U \neq 0$
can leave the spacetime $a=0$ irrevocably. Objects with the
``three-dimensional'' energies $E$ being -- instead of real numbers --
imaginary ones, behave in a similar way. Note that the above condition
does not have to be fulfilled by the representation (\ref{waver}).
Depending on the sign of the imaginary component of the quantity $E$,
it could either approach zero or amplify illimitably with the lapse of
time $t$. Such a behaviour of the representation (\ref{waver}) is not
surprising, as the equations of quantum mechanics do not take into
account the existence of objects with complex energies $E$. Thus, the
above discussion clearly corresponds to the {\it a priori} requirement
of quantum mechanics that demands the Hamiltonian of an object or, more
generally, of a quantum system to be a Hermitian operator, i.e.\ to have
eigenvalues in real numbers, since those eigenvalues correspond to
real, i.e.\ measurable physical quantites.\footnote{Note, however, that
for instance the so-called quasi-stationary states are formally
described in quantum mechanics with the use of the complex energy $E$,
whose imaginary part has a negative value.} And indeed, according to
the toy model, the lack of fulfillment of this requirement --
which means that the energy $E$ of an object has not a real value --
results in the escape of a given object from the space $a = 0$.  For
instance, such a situation could occur for an ``exotic'' kind of
particles which are tachyons. Provided that the tachyonic energy $E$
fulfils the condition $E^2 < 0$, we then have $\omega^2 < 0$, so for
the values of $U$ such that $U^2 > E^2$ one obtains 
$a ( \eta ) \sim \pm {\rm sinh} (|\omega| \eta)$ and the tachyon would 
disappear in the additional spatial dimension $a$.

Now let us consider the equality (\ref{r2}). Assuming that $p=0$ 
and hence $E = m c^2 = \hbar\omega$, we obtain a formula describing 
the motion of a particle that
appears in the ``observable'' zone $a \in [-L, L]$ periodically but each
time for a very short, comparable to $T_l$, time only.  Such particles
may correspond in our toy model to the actual phenomenon of the 
zero-point energy. Note that these
particles behave exactly in the way the Planck oscillators do.  
We can also state that the toy model permits one
to pose the rarely asked question beginning with the word {\it why?}
Namely why, in -- or with respect to -- any (inertial) coordinate
system in the spacetime $a=0$, massless particles have always the same
velocity, equal to the speed of light $c$? The toy model provides an
opportunity to answer this question. Let us imagine the massless
particle, whose velocity is not constant in every coordinate system of
the spacetime $a=0$. Then one can find such a coordinate system in
which the particle velocity is equal to zero in the three-dimensional
macrospace $a = 0$, so the particle seems to remain motionless if
only the macrospace $a = 0$ is taken into account.  However, in this 
coordinate 
system one has $a=\pm U \gamma$, and in effect the particle leaves the
spacetime $a=0$ irrevocably, never crossing it again.  In another
coordinate system of the spacetime $a=0$, in which the particle is
moving, we obtain $a \sim \pm \sin (\omega \gamma)$.  In such a case, 
the particle remains in the ``neighbourhood'' of the spacetime $a=0$,
perpetually crossing the hypersurface $a=0$.  Thus we arrive at the
evident contradiction, as in one coordinate system of the spacetime
$a=0$ the number of crossing of the hypersurface $a=0$ is equal to zero,
whereas in another system it equals a certain natural number which
grows with increasing the parameter $\gamma$. Thus, we reach the
obvious conclusion that the constant velocity of massless
particles in all coordinate systems of the spacetime $a=0$ ensures
the self-consistency of the toy model.

It is also worth mentioning another important feature of the solutions
(\ref{r1}) and (\ref{r2}).  Namely, according to Eqs.~(\ref{e1}) and
(\ref{e2}), the arguments of the sine functions in these expressions
can have a minus sign, corresponding to particles with negative
energies $E$, i.e.\ antiparticles. They would then be formally
equivalent -- if taking into account their motion with respect to the
dimension $a$ -- to the ``ordinary'' 
particles with positive energies $E$, but with the symmetrically reversed 
($\eta \to - \eta$) lapse of the affine parameter, or the proper time, 
$\gamma$ or $\eta$.  Taking into account other, ``macroscopic'' 
spatial dimensions $x$, $y$ and $z$, the antiparticle
would then be nothing but the particle moving -- with respect to the
spacetime $a=0$ -- in the opposite direction to that of the motion
of an ``ordinary'' particle. Note as well that, according to the
formulae (\ref{r1}) and/or (\ref{r2}) and to the relationship $\sin
(-x) = - \sin x$, the antiparticle would also be formally equivalent to
the ``ordinary'' particle with the same, as for the antiparticle, lapse
of the affine parameter $\gamma$ or $\eta$, but moving in the opposite
direction (only) with respect to the dimension $a$.

Now we will draw some attention to the quantity $U$. It is the total
energy of a particle, which remains constant for particles travelling
along the geodesic line, since the metric (\ref{2}) is independent of
time. A condition, which should be satisfied by the energy $U$ if the
particle's trajectory is to be time-like or null (as calculated with
respect to the dimension $a$), can easily be derived,
\begin{equation}
U^2 \le \hbar^2 |\lambda| c^2 + E^2 \ .
\label{condition}
\end{equation}
Hence, the maximum depth the particle can penetrate into the additional
spatial dimension is equal to
\begin{equation}
a_{\it max} = \pm \frac{\hbar c}{E} \ .
\label{d_3}
\end{equation}
In particular, we obtain $a_{\it max} \to \pm \infty$ for $E \to 0$.

So far, within the framework of our model, it is difficult to add 
anything concerning
the quantity $U$. Let us imagine, however, a particle with a total
energy $U$, an energy $E$ in the three-dimensional space $a=0$, and 
a frequency $\omega$. The particle moves with a uniform rectilinear
motion with respect to the spacetime $a=0$ and simultaneously
oscillates in the additional spatial dimension $a$.  Then, let us
assume that this particle encounters a potential barrier $V > E$ on its
way; the potential is relatively shallow with respect to the additional
spatial dimension; see Fig.~1. It is evident that an ordinary
``classical'' particle, moving only in the four-dimensional spacetime
$a=0$, is not able to surmount the potential barrier higher than the
particle's total energy $E$. When an additional spatial dimension is
present, however, then the situation changes substantially.  If the
potential $V$ is very shallow with respect to the extra spatial dimension, 
then the particle can circumvent it (see Fig.~1), even when its
total energy $U$ does not exceed $E$ by much. There is nothing puzzling
about this phenomenon; on the contrary, its simplicity -- which is the
result of purely geometric relationships between the particle's motion
and the spacetime configuration -- is striking.  The particle vanishes
from ``our'' spacetime, given by $a=0$, just in front of the barrier
to reappear soon after circumventing it. There is obviously a certain
possibility that the particle might be reflected from the barrier. It
is clear that the transmission (or reflection) coefficient depends on
the particle's energy $E$, so -- consequently -- on its oscillation
frequency $\omega$ as well as on parameters characterizing the
potential barrier. In such a case the ``hidden''
(but inherent) parameter $U$ can also be a quantity that significantly
influences the particle's behaviour.\footnote{On the other hand, let us
note that -- similarly as in the ``standard'' quantum mechanics -- 
many of the measurable quantities associated with particles (or waves),
according to the formulae (\ref{r1}) and (\ref{r2}), seem to be 
independent of some of the particle/wave parameters, 
such as (for instance) its oscillation amplitude, and hence -- 
also of the energy $U$.} It is difficult to say
what laws govern this parameter in a multi-particle system:\ is
it characterized by some probability distribution, in the common sense
of this notion coming from statistical physics?  If it is, what does
such a probability distribution
depend on? One expects that in the case of more ``realistic''
particles than those just discussed (i.e.\ possessing spin, charge, {\it
etc.}), in their description there would also appear ``hidden''
parameters other than $U$, characterizing for instance the total
four-dimensional spin and/or charge of the object, 
or of the system of objects. It is
possible that ``hidden'' parameters may be responsible for subtle
correlations between distant objects which were previously interacting.

Note that the fact of the
existence of the ``hidden'' parameters -- such as the energy $U$, or
the four-dimensional ``super-spin'' and/or ``super-charge'' 
of the system -- indicates that the
toy model is a {\it non-local theory} which contains {\it hidden
variables}. Thus, the toy model -- fulfilling the
Bell inequality -- simultaneously could reconcile the
Einstein--Podolsky--Rosen paradox \cite{EPR}. We should,
however, point out that our model's particles remain ``classical'',
irrespectively of our knowledge of ``hidden'' parameters and the values
which they take.
On the other hand, sets of particles within
the framework of the toy model can be subject to a statistical
interpretation. Namely, two objects can exist in two different states
even if they have the same values of ``standard'' three-dimensional
parameters. However, they can still vary in values of ``hidden''
parameters, such as (for instance) the energy $U$.  Perhaps, in the case 
of a multi-particle system one can speak of a probability distribution,
characterizing those ``hidden'' parameters.  However, rules controlling
such distributions as well as possible reciprocal (mutual) variations 
of ``hidden'' parameters at the moment of interaction of objects which 
are characterized by those parameters, remain so far unknown.

It then seems that in the toy model we may interpret the probability
rules of quantum mechanics as statistical results of a behaviour (or
changes) of completely determined values of variables which are hidden
to us. One may illustrate that on a simple 
example:\ the probability of finding a (non-relativistic)
``quantum'' particle within some region of the configuration space is
connected to the function $|D|^2$ integrated over the volume of this
region.  In general, the quantity $D$ denotes here the (normalized)
wave-function representing the particle.  However, in the case of a
particle encountering potential(s) which are all time-independent, it
is sufficient to assume that the quantity $D$ is the (normalized)
spatial part of the wave-function representing this particle; see
expression (\ref{waver}) which represents the simplest case of 
a wave-function -- that for a free particle, 
i.e.\ a wave-function in the absence of
any potentials.  The toy model adds some new elements to the above
picture coming from the ``standard'' (non-relativistic) quantum mechanics.
Namely, the particle's wave-function occurring in quantum mechanics
corresponds in the toy model to the function describing the particle's
motion, i.e.\ also its oscillations in the additional dimension of
spacetime. In turn, the function characterizing the motion of the
toy-model particle depends, among others, on the value of the
``hidden'' parameter $U$ which is the particle's total energy in the
five-dimensional spacetime; see the factors preceding the sine
function in the formulae (\ref{r1}) and (\ref{r2}) for the simplest
case of motion -- that of a free particle. It is clear that two different
objects which have identical values of all their {\it directly
observable} parameters (such as the oscillation frequency $\omega$), 
may possess different values of the {\it hidden}, but {\it inherent}
parameter $U$. Furthermore, one may also imagine a set of such apparently
``identical'' (non-interacting) particles which may have different (but
always ``completely'' determined) values of the quantity $U$. In the
case of a large set of such particles in the configuration space, it
seems that the values of their ``hidden'' energies $U$ may be
characterized by some probability distribution;\footnote{It seems 
that in the considered case the probability distribution of 
the parameter $U$ could
be estimated in a usual manner; for instance, one might imagine
constructing a histogram:\ $n_1$ particles from the set take the values
of the parameter $U$ from the interval $[U_1, U_1 + \Delta U]$, $n_2$
particles -- from the interval $[U_2, U_2 + \Delta U]$, $\ldots$, and
finally $n_N$ particles -- from the interval $[U_N, U_N + \Delta U]$.
Then, the formal limits $\Delta U \to 0$ and $N \to \infty$ remain to
be performed and the probability $p = p(U) \equiv n/N (U)$ to be
calculated.} this probability distribution should then determine the
value of the ``hidden'' energy $U$ of every single particle from
the above-mentioned set of particles. Thus, the distribution of
probability of finding a particle within some region of the
configuration space, connected to the quantity $|D|^2$ in
(non-relativistic) quantum mechanics, seems simply to result from, or
to be determined by the probability distribution of the inherent
parameter $U$.

\bigskip

Now, for the moment, let us consider a wave motion in general. 
From the Fourier analysis one knows that if the ``dispersional'' 
extent of the wave group in the $i$-th dimension
equals $\Delta x_i$ and the indeterminacy, or ``bandwidth''
of the wave-vector in this
particular dimension equals $\Delta k_i$, then both these quantities
fulfil the following inequality
\begin{equation}
\Delta x_i \, \Delta k_i \ge \frac{1}{2} \ ;
\label{h2}
\end{equation}
it occurs for $i = 2, 3, 4$.
Likewise, if the impulse duration comprises the time interval $\Delta t$
and the frequency of the group is ``spread over'' the range $\Delta
\omega$, then both these quantities fulfil the relationship
\begin{equation}
\Delta t \, \Delta \omega \ge \frac{1}{2} \ .
\label{h3}
\end{equation}

Let us now return to the toy model and assume that all objects, whose
behaviour we investigate here, move along their geodesic lines. We limit
our considerations to the case of a given wave, or of a particle whose
trajectory oscillates in the extra spatial dimension as well as to the
case of a set of ordinary particles, i.e.\ those with negligible
gravitational ``interactions'' among them.  For the trajectory 
of the investigated object, we may write $\Delta {\bf
p} = \hbar \Delta {\bf k}$ and $\Delta E = \hbar \Delta \omega$ where
the quantities $\Delta {\bf p}$ and $\Delta E$ are defined analogically
as the quantities $\Delta {\bf k}$ and $\Delta \omega$ above, with
taking into account the relationships ${\bf p} = \hbar {\bf k}$ and
$E = \hbar \omega$ which hold for each of the monochromatic components
of the particle's motion trajectory (of course, some of those
monochromatic components may be ``shifted'' with respect to the
spacetime $a = 0$ and centred at some $a \neq 0$). 
Hence, according to expressions
(\ref{h2}) and (\ref{h3}), we have
\begin{equation}
\Delta x_i \, \Delta p_i \ge \frac{\hbar}{2}
\qquad \mbox{for $\;$ $i = 2, 3, 4$}
\label{h4}
\end{equation}
and
\begin{equation}
\Delta t \, \Delta E \ge \frac{\hbar}{2} \ .
\label{h5}
\end{equation}
We can then assume that for a time interval $\Delta t$ sufficiently
long, one has $\Delta E \cong 0$, or $E \cong {\it const}$ for our
particles.\footnote{In other words, in order to make sure that the
energy $E$ of the oscillating particle is the quantity remaining almost
constant in time, one should observe the particle's motion for a
sufficiently long (or ``almost infinitely''long) interval of time
$\Delta t$.} Note that, from the standpoint of the considered model,
there is nothing strange about the relationships (\ref{h4}) and
(\ref{h5}).  On the contrary, they are consistent with the dynamics of
the particle's motion. For instance, as we were able to see earlier while
analysing expression (\ref{r2}), the particle 
with the energy $E=0$ would move
according to the formula $a=\pm U \eta$, instead of remaining motionless
in the spacetime $a=0$. In effect, such a particle would irrevocably
vanish from the spacetime $a = 0$, without violating the conditions
(\ref{h4}) and (\ref{h5}) which hold for the particle oscillating in
the additional spatial dimension, i.e.\ for the one that periodically
appears in the spacetime $a=0$.  Moreover, one should note that, in
our case, expressions (\ref{h4}) and (\ref{h5}) refer to the single
particle travelling -- in the case of the metric (\ref{2}) -- with the
harmonic motion, or moving in the presence of some potentials, as well
as to a multi-particle system.

\bigskip

Let us now imagine performing such an experiment: Two plates are
inserted in a very-low-temperature gas so that the noise and thermal
motions of gas particles are minimized. We might imagine that the
plates remain almost at rest in the spacetime $a=0$.(\footnote{For a
macroscopic object one has $a \approx 0$, as the object consists of a
large number of particles, between which the gravitational
``interactions'' are negligible, 
so the superposition of all the particles'
oscillations is close to zero.}) Moving gas particles oscillate in the
spatial dimension $a$ according to the formula (\ref{r2}), so between 
the two plates, only the particles which 
cross the hypersurface $a=0$ at the places
$(a \approx 0, x, y, z)$ where the plates are situated, can rebound
from the plates and exist between them. Obviously, these would be the
particles for which the quantity -- half of a period of oscillations
times the velocity with respect to the spacetime $a=0$ -- fits the
distance between the plates in natural numbers.  Other particles simply
circumvent the plate(s) in the space $a \ne 0$; see Fig.~1. Outside the
plates all modes of the particles' oscillations are possible. This means
that fewer particles will be found between than outside of the plates,
so a net pressure will drive the plates together; and this seems to be
nothing else but the Casimir effect.

In similar ways, one can analyse -- with respect to our particles --
diffraction, scattering, tunnelling into the classically forbidden
region and other phenomena.  The conclusion is quite clear -- the toy
particle, i.e.\ a purely classical relativistic particle moving along
the geodesic line in the five-dimensional spacetime of the toy model,
behaves in a way very similar to the quantum particle. However, in 
contrast to the real quantum particle, the toy particle's behaviour is
very easy to explain, even with simple pictures; see, for instance,
Fig.~1. This is because, in our case, the particle's behaviour is merely
dictated by the geometry of the spacetime, in which the particle is
moving. The conclusion of this section can then be expressed as
follows:\ in our toy model, the quantum effects actually result from a
purely classical structure of the ``microspace''.  Quantum-mechanical
effects, present in the toy model, are then the measurable consequence 
of the existence of the additional spatial dimension which is
very hard to be detected directly.  In other words, it seems that the 
entirely classical theory, which describes some phenomena occurring in 
the five-dimensional spacetime of the toy model, results in the theory 
of the quantum mechanics when one confines oneself to considering those 
phenomena as if they took place exclusively in the four-dimensional 
spacetime given by $a = 0$; in this section we have seen that
such an approach leads, among others, to the description of some purely 
corpuscular phenomena with the use of the wave-formalism.

Thus, equation (\ref{1}) describes in the toy model a possible widest 
class of phenomena, ranging from the microworld to the entire Universe. 
The remaining question is whether we could allow ourselves to abandon 
the equations of quantum mechanics in the discussed model. It seems 
that we could not. Indeed, the problem of the quantum behaviour of 
a particle in a given situation might be settled by
finding a solution to Eq.~(\ref{1}) with an appropriate stress--energy
tensor.  However, it is easy to imagine how difficult and arduous 
a task this would be. Therefore, a much more practical approach is to 
retain the Schr\"odinger (or rather the Klein--Gordon) 
or the Dirac equations, bearing in
mind that a comprehensive and {\it fully relativistic} handling of a
``quantum'' issue would be provided by solving Eq.~(\ref{1}) with an
appropriate stress--energy tensor. Thus, the Klein--Gordon or the Dirac 
equations can be, in the toy model, regarded as a kind of approach to 
the Einstein generalized equation (\ref{1}) with a stress--energy 
tensor suitable for the investigated problem.

At the end of this section, let us note that ordinary particles (with
finite masses) as well as massless ones should experience similar
effects and phenomena.  It is a consequence of the wave character of
motion of both the particle types, which motion is described in both 
cases by very similar mathematical formalism; see the formulae (\ref{r1}) 
and (\ref{r2}). And indeed, there exist optical phenomena which are the
exact analogues of each of the quantum effects related to the motion of 
a free particle with a finite mass.

\bigskip

Among numerous questions that arise as a result of the analysis of the
particle's motion in the spacetime given by the metric (\ref{2}), a few
come to the fore:

{\bf i)} By what laws is the quantity $U$ governed? Does this quantity
have any probability distribution in the multi-particle system? If so,
do statistical distributions known from quantum mechanics reflect the
distribution of the quantity $U$?

{\bf ii)} Since, in our model, the quantum physics and its laws were 
not introduced directly, then what actually should be understood under 
the term ``particle'' in the toy model? In this situation, the only
reasonable answer seems to be the model of a particle as a specific
entity ``formed from'' the classical fields.

{\bf iii)} What is the shape of ``average'' potentials with respect to
the extra spatial dimension $a$?  Is this shape similar for different
potentials or is it dependent upon the potential type? Perhaps the
shape of a potential with respect to the dimension $a$ is influenced by
the spacetime structure in such a way that the shape is formed (or
``forms itself'') as a certain state of the dynamic equilibrium between
the potential and the spacetime?

{\bf iv)} How could charged particles be introduced into the discussed
toy model or, in general, how could the electromagnetic interactions,
as well as other interactions -- weak or strong, be taken into account
within the framework of this model?\footnote{Maybe, the first -- and
rather phenomenological -- step in solving this task could consist in
introducing into the toy model the five-dimensional wave equation and
in investigating its properties; see Refs.~\cite{Souriau,Unwin}.} Does
their presence result from specific spacetime deformations? Perhaps
the answer to the last question is negative and therefore the
electromagnetic, strong and weak interactions are transferred by the
boson force carriers, instead of spacetime deformations.

In general, one may imagine that there exist two different types of
solutions to the formally identical equation (\ref{1}). The first type
of solutions, represented by the metric tensor $g_{\mu\nu}$, would
describe usual gravitational ``interactions'' which result from the
spacetime deformations caused by the spacetime distribution of
matter/energy.  This type of solutions is actually investigated in the
present paper and composes the contents of the toy model.  The second
type of solutions to equation (\ref{1}) would correspond to all kinds
of ``typical'' (i.e.\ not gravitational) interactions:\ the strong, weak,
and electromagnetic ones (and, maybe, others -- not detected so far).
These Kaluza--Klein-type solutions would be in the form of a
``potential'' tensor (i.e.\ in the form of a tensor of ``potentials'')
with components characterizing the above-mentioned interactions, such as
the four-vector potential $A_\mu$ for the electromagnetic forces. 
The second type
of solutions might possibly be subject to some quantization-like
procedure(s). Maybe, there also exists the third type of solutions to
the Einstein equation (\ref{1}), which would describe the (internal)
structure of elementary particles.

\section{Schwarzschild metric}

The aim of this section is to check whether the toy model introduces
any corrections to the predictions of the ``standard'' general relativity
regarding, among others, the so-called ``five tests of general
relativity''; see Ref.~\cite{Wheeler}.  
All these tests are a consequence of a particular
solution to the Einstein original equation, known as the Schwarzschild
metric. Let us consider how this metric will be affected by the
introduction of the extra spatial dimension.

In the presence of a spherically symmetric object with a mass $M$ in an
empty (asymptotically flat) space, an exact solution of equation
(\ref{1}) with the stress--energy tensor given by expression
(\ref{t}) takes the following form,
\begin{eqnarray}
ds^2 = \left(1 - \frac{2 m_M}{r} \right)\!  \left(1 + |\lambda|
a^2\right)\!  c^2 dt^2 - \left( 1 + |\lambda| a^2\right)^{\! -1} da^2
\nonumber\\
- \left( 1 - \frac{2 m_M}{r} \right)^{\! -1} d r^2 - r^2 d \theta^2 -
r^2 \sin^2 \theta \, d \varphi^2 \ ,
\label{schw}
\end{eqnarray}  
where $r\in (0, \infty)$, $\theta\in [0, \pi]$, and $\varphi\in [0,
2\pi)$ are the polar coordinates of the ``macroscopic'' three-dimensional
space around the mass $M$, and $m_M \equiv GM/c^2$.  In the field
of a spherically symmetric object with the mass $M$, a solution
to the equation of motion of a test particle with respect to the 
dimension $a$ can be written as
\begin{equation}
a(\eta) = \pm \sqrt{\frac{U^2 - E^2}{E^2 |\lambda|}} \,
\sin (\omega \, \eta)
\label{solsch}
\end{equation}
where
\begin{equation}
\omega \equiv \frac{E}{\hbar} \!\left( 1 - \frac{2 m_M}{r} 
\right)^{\! -1/2} \ , 
\label{om*}
\end{equation}
the quantity $\eta$ is an affine parameter, and $E$ denotes 
the particle's ``three-dimensional'' energy, defined exactly
in the same form as in the case of the Schwarzschild original solution;
then -- for instance -- for a particle with a finite (non-zero) mass $m$
one has
\begin{eqnarray}
E^2 = c^2 \!\left( 1 - \frac{2 m_M}{r} \right) \!\left[ m^2 c^2 \right.
\nonumber \\ \left.
+ \left( 1 - \frac{2 m_M}{r} \right)^{\! -1}
\!\left(p^r\right)^2
+ r^2 \!\left( p^\theta \right)^2 + r^2 \!\left(\sin^2 \theta \right)
\!\left( p^\varphi \right)^2\right]
\label{s_energy}
\end{eqnarray}
where $p^r \equiv m \, dr/d\tau$, $p^\theta \equiv m \, d\theta/d\tau$,
$p^\varphi \equiv m \, d\varphi/d\tau$, and $\tau$ denotes the proper
time of the particle whose motion is under consideration.  Recall from
the previous section that, for sufficiently large time intervals, the
quantity $E$ remains almost constant for a particle moving along 
the geodesic line.

Below, it is assumed that the mass $M$ of the investigated object or
system of objects is small when compared with its size:\ $2 G M / 
(c^2 r) \ll 1$.

\subsection{Five tests of general relativity} 

{\bf Gravitational redshift.} The general theory of relativity predicts
that the spectrum of the radiation 
emitted in the neighbourhood of a massive
object and receding from it, should be shifted towards the red part of
the spectrum. In turn, the spectrum of light moving in the direction of
such an object should show shift towards the violet part of the
spectrum.

Let us suppose that an electromagnetic wave or a wave of matter has a
frequency $\omega_1$ in the distance $r_1$ from the mass $M$.  In
the distance $r_2$ from this object the wave frequency is $\omega_2$.
Then, according to the formula (\ref{om*}), the gravitational redshift
defined as $z \equiv (\omega_1 - \omega_2)/\omega_2$ is given by
\begin{equation}
z = \left( \frac{1 - 2 m_M/r_2}{1 - 2 m_M/r_1} \right)^{\! 1/2} - 1
\cong m_M
\frac{r_2 - r_1}{r_1 r_2} \approx \frac{gh}{c^2} \ ,
\label{shift}
\end{equation}
where $g\equiv GM/r^2$ (with some $r$, outside the mass $M$, such that
$r_1 \approx r \approx r_2$) and $h \equiv
r_2 - r_1$.  The above formula is identical to that obtained on the
basis of the considerations concerning the energy conservation for a
particle (or wave) moving along the geodesic line in the
four-dimensional $(1+3)$ spacetime of the ``standard'' general relativity,
described by the Schwarzschild original metric.  It should be stressed,
however, that in the standard theory the explicit relationship -- such 
as the formula (\ref{om*}) in the toy model -- 
between the frequency $\omega$ of
the particle's oscillations and the ``field'' factor $(1 - 2 m_M/r)$ does
not exist. It happens, since there is no direct connection between the
quantity $\omega$ and the spacetime curvature in the ``standard'' general
relativity.

\bigskip

The remaining well-known tests of general relativity concern propagation
delay of radar signals, light deflection, perihelion precession
and geodesic precession. After some calculations one can state 
that the values of
all measurable parameters, characteristic for the above-mentioned
processes or phenomena, derived within the framework of the toy model
are the same as those obtained for the case of the
Schwarzschild original metric within the ``standard'' general 
relativity.

\subsection{Dark matter}

In this section, we argue that there are no reasons for
postulating the existence of dark objects. We also try to show that
effects which are usually attributed to the presence of a hypothetical
non-luminous matter can be explained readily by {\it geometric}
properties of a spacetime which contains an additional spatial
dimension denoted by $a$.(\footnote{It is worth adding that some 
authors have suggested solutions to the
missing material problem, by the manifest changing of 
Einstein's law of gravity \cite{gal,pr,pr1}.})

Presumably the most ``spectacular'' need for a dark material is provided
by the observational data concerning spiral galaxies. These objects
seem to have a constant circular rotation velocity $v_c$, independently
of the galaxy radius $R$ (apart from the relatively small, initial
values of $R$ when, obviously, $v_c$ approaches zero as $R \to 0$).  On
the other hand, the brightness of matter -- and the density of a luminous
material, if we reasonably assume that the brightness-to-density ratio
remains constant -- in spiral galaxies decreases exponentially with
growing radius $R$. Simple calculations based on the Newtonian mechanics
suggest, however, that for a constant circular velocity $v_c$ the density
of matter should decrease as $1/R^2$, i.e.\ much more slowly than
exponentially \cite{Peebles}. 
One therefore assumes that, in addition to the ``ordinary'' visible
matter, there also exists a dark material that causes the
observational data to remain in agreement with reasonable (for example,
on a scale of the solar system) predictions of the Newtonian mechanics.
The dark matter should be concentrated in outer parts of a galaxy, 
forming the so-called dark halo, as the Newtonian calculations and 
astronomical observations give practically identical results only 
for small (initial) values of the galaxy radius $R$. 
Such a situation seems to be a general feature of all the known 
spiral galaxies. This is, given the dark matter, somewhat weird, 
since both the luminous and the dark material
should possess identical gravitational properties. Thus, both kinds of
the material should display -- independently of one another -- rather
random spatial distribution in different observed galaxies. Therefore,
for various galaxies a completely different dark matter distribution
inside the galactic space might be expected.

There are also other reasons to postulate 
the existence of dark matter \cite{Peebles,gal}. 
In this paper, however, we will focus on probably the most spectacular
case, i.e.\ that of spiral galaxies.  Other cases such as those
concerning the motion of so-called dwarf galaxies \cite{Peebles} as
well as of regular clusters of galaxies \cite{bohr1,bohr2} which are 
in a virial equilibrium, may be approached in a very similar manner.

For the purpose of studying large astronomical systems, such as
spiral galaxies, usually the so-called virial theorem 
is being applied \cite{gal}. In the case of the Newtonian mechanics, 
the virial theorem gives
\begin{equation}
\frac{GM}{R} = 2 \sigma^2
\label{dm1}
\end{equation}
where $M = M(R)$ is the mass contained within the gravitational 
radius $R$ and $\sigma$ denotes the velocity
dispersion; see, for instance, Refs.~\cite{Peebles,gal}.  
Note that for a spiral galaxy one has $v_c = 2^{1/2} \sigma$.
As it was pointed
out earlier, equation (\ref{dm1}) implies -- if $v_c (R) = 
{\it const}$ is assumed -- the spatial distribution of the mass of a
spiral galaxy which is inconsistent with observations. Now let us
note that the virial theorem applied in the toy model assumes a
form slightly different from Eq.~(\ref{dm1}), namely
\begin{equation}
\frac{GM}{R} + \frac{c^2}{2} |\lambda| a^2
= 2 \sigma^2 \ ,
\label{dm2}
\end{equation}
for $|\lambda| a^2 \ll 1$. Due to the rapid oscillations of the
quantity $|\lambda| a^2$ in time, one will detect an ``expected'',
i.e.\ mean (or average) value of this quantity over time. Thus, in the
formula (\ref{dm2}), the quantity $|\lambda| a^2$ should rather be
substituted by its average value,
\begin{equation}
\left\langle |\lambda| a^2 \right\rangle_\tau 
= \frac{U^2 - E^2}{2 E^2} \equiv U_E \ .
\label{avera}
\end{equation}
Note that for a value of the energy $E$ close to $U$, we have $U_E
\approx (U - E)/E$.

Asking whether the value of the quantity $U_E$ can depart considerably
from zero for a large macroscopic object, such as a planet or even a
part of a galaxy, should start the analysis of Eq.~(\ref{dm2}).  In
order to answer this question, let us return for a moment to the formulae
(\ref{r2}) or (\ref{solsch}).  Note that the ``classical'' motion
of a particle in the four-dimensional spacetime $a=0$ of the toy model
corresponds in the discussed case to the performing of a formal limit
$|\lambda| \to \infty$.  Note also that this operation is equivalent to
taking of the limit $\hbar \to 0$, as one has $|\lambda| \propto
\hbar^{-1}$. Thus,
within the ``classical'' limit we obtain $|\lambda| \to \infty$, and
consequently $a(\eta) \to 0$. However, the product $|\lambda| a^2$ can
remain a finite, non-zero number.

Let us recall expression (\ref{avera}); we see that in the case of 
studying the virial theorem in the context of dark matter, there again 
arises the problem of the meaning of the total energy $U$.  The first
question that emerges is whether the quantity $U_E$ depends on
parameters -- such as the mass -- of the investigated object and, 
if so, how it depends on it.  It seems possible that $U_E$ could take
relatively large values for ``quantum'' particles, i.e.\ for particles 
which have relatively low masses and/or energies $E$ 
[see the formula (\ref{d_3})], and very small values for ``classical'' 
massive objects (see footnote 11 in section 3.1), although this does 
not have to always be true and remain a rule.

\bigskip

Now let us make a digression referring to the mass amount. In
general, the maximum acceptable mass of a single particle seems to be
the Planck mass $m_{\it Pl} \equiv h/(c^2 T_l)$; see section 3 and
the formula (\ref{4}) therein. Note that an object with the Planck
energy $E_{\it Pl} \equiv m_{\it Pl} c^2$, 
in the anti-de Sitter spacetime
${\cal S}^1 \times {\cal R}^1$ whose covering surface is given by the
metric (\ref{3}), oscillates in this spacetime with a vibration period
equal to the Planck time $T_l = h/E_{\it Pl}$.  A very interesting
relationship can be observed:\ namely, objects from this anti-de Sitter
spacetime, which have vibration periods equal to $n T_l$, where $n \in
{\cal N}$, retrace their own life histories exactly each time after the
lapse of the coordinate time $n T_l$. This relationship -- 
which is a condition of the
particles' time-like (geodesic) curves closure in the anti-de Sitter
spacetime ${\cal S}^1 \times {\cal R}^1$ -- could be called the
resonance condition of the particles' vibration in the anti-de Sitter
spacetime.  Note that in the discussed phenomenon, the Bohr--Sommerfeld
quantum condition(s) are satisfied, as we have
\begin{equation}
\oint p_q \, dq = \oint \frac{h}{\Lambda} \, d (ct) = 
\frac{h}{c \, n_q T_l} c \, n_q T_l = h
\label{oint}
\end{equation}
where the quantities $\Lambda$ and $c$ denote, 
respectively, the wavelength and the
velocity of the oscillating particle with respect to the hypersurface
$a = 0$, i.e.\ in the $(ct)$-direction of the anti-de Sitter spacetime.

\bigskip

Let us now turn to our original problem of dark matter and assume that
the distribution of the mass density is consistent with observational
data, i.e.\ it exponentially decreases with growing galaxy radius $R$.
For sufficiently large distances $R$ from the spiral galaxy centre,
a component $GM/R$ on the left-hand side of Eq.~(\ref{dm2}) takes a
(relatively) small value, so we assume that it becomes rather
negligible as compared with $c^2 \langle |\lambda| a^2 \rangle_\tau/2$.
On the other hand, for distances $R$ sufficiently large, the relative
growth in the mass of the galaxy $\Delta M/M$ within the radius $R$,
along with increasing $R$ by $\Delta R$, is relatively small. In
such a case, also the value of the relative increase in the momentum
$\Delta p/p$ -- along with the growth of $R$ by $\Delta R$ -- is
comparatively small, as the circular velocity of the galaxy is close to
constant for large values of $R$.  From this it immediately follows 
that the corresponding relative increments in the energies $\Delta E/E$
and -- as a consequence of all that was said above -- also $\Delta U/U$
are comparatively small, too;\footnote{It seems to be rather improbable,
also from the statistical point of view, that the ``hidden'' energy
$\Delta U$ of the mass $\Delta M$ of the outer part of the galaxy --
for which $\Delta M/M \approx 0$ and $\Delta E/E \approx 0$ --
does not fulfil the condition $\Delta U \ll U$.} see expressions 
(\ref{solsch}) and (\ref{s_energy}), where for the considered case
one has $p^r = 0 = p^\theta$ and the circular velocity $v_c$ is given by
the relationship $v_c = r (\sin \theta) p^\varphi/m$.  Therefore we can
assume -- on the basis of the relation (\ref{avera}) -- that for
relatively large distances $R$, we have $\langle |\lambda| a^2
\rangle_\tau \approx 
{\it const}$. Thus, there holds
\begin{equation}
\frac{c^2}{2} \!\left\langle |\lambda| a^2 \right\rangle_\tau
\approx v_c^2 (R) \approx {\it const} \ .
\label{dmc}
\end{equation}
For a typical spiral galaxy, the constant circular velocity is of 
the order of $10^2$ km s$^{-1}$.  Then
\begin{equation}
\left\langle |\lambda| a^2 \right\rangle_\tau = U_E \sim 10^{-7}\ ,
\end{equation}
which seems to be a reasonable result considering that the discussed
system seems to be highly ``classical''.

Now we will examine what happens within the small distance from the
galaxy centre. Assuming that $R \to 0$ we obtain $|\lambda| a^2 \to 0$,
as for $R \to 0 $ the relation $U \to E \to 0$ takes place; see the 
formula (\ref{solsch}). It is because for $R = 0$ obviously holds $U =
E = 0$ and $a(\eta) = 0$ (as there is no mass right in the middle {\it
point} of a spiral galaxy, since the topological measure of a point in
a three-dimensional space or of a one-dimensional line in a
four-dimensional space is equal to zero), whereas we do not expect the
function $a(R)$ to be not continuous.  It seems that the quantity
$\langle |\lambda| a^2 \rangle_\tau$ is a monotonic function of $R$,
reaching plateau for sufficiently large values of $R$, for which --
along with the further growth of $R$ -- the value of the quotient $U/E$
remains virtually on the same level, at least as compared with the rate
of the earlier growth.

Let us note that due to the presence of the ``hidden'' parameter $U$ of
an unknown distribution $U = U (M, R)$ in the quantity $\langle
|\lambda| a^2 \rangle_\tau$, theoretical results can in principle be
perfectly fitted to experimental data. In other words, the comparison
between observational results and Eq.~(\ref{dm2}) might allow us to
determine the function $U = U(M, R)$. One of many possibilities of a
very precise representation of experimental data by Eq.~(\ref{dm2}) is
provided by the following relationship
\begin{equation}
\left\langle |\lambda| a^2 \right\rangle_\tau = \alpha R
\label{dm3}
\end{equation}
where $\alpha$ seems to be a ``universal'' constant, the value of
which, $\alpha \sim 10^{-28}$ m$^{-1}$, allows one to reproduce to 
some extent the experimental data for numerous large astronomical 
systems of various
types and sizes.\footnote{Detailed investigations of a number of spiral
galaxies, with taking into account a linear contribution to the
potential function, were performed by Mannheim \cite{Mann1,Mann2};
note that in Ref.~\cite{Mann2} one assumes that $\alpha \equiv \alpha_1
+ N \alpha_2$ where $\alpha_1$ and $\alpha_2$ are universal constants
and $N$ denotes the total amount of visible -- stellar (and gaseous) --
matter in solar mass units in a galaxy. The
presence of the potential given by the right-hand side of expression
(\ref{dm3}) in the virial equation (\ref{dm2}) also permits one to
reproduce the experimental data concerning both regular clusters of
galaxies and dwarf spheroidal galaxies.} Let us note that the
potential given by the formula (\ref{dm3}) evidently does not satisfy the
relationship (\ref{dmc}), as for expression (\ref{dm3}) one has $U^2 =
E^2 (1 + 2 \alpha R)$. The existence of such a relationship, however,
seems to be far less possible than of expression (\ref{dmc}) which 
implies that $U/E \approx {\it const}$ 
for sufficiently large values of the quantity $R$.

In a similar manner to that described above we can explain phenomena
attributed to the presence of dark matter in astronomical objects other
than spiral galaxies, e.g.\ in so-called regular clusters of 
galaxies as well as in dwarf 
spheroidal galaxies \cite{Peebles}. Hence, the
presence of the extra spatial dimension enables the explanation of the
dynamics of large astronomical objects without introducing a substantial
amount of any dark matter. Of course, it does not rule out possibility
that a part of the matter in the investigated objects exists in a
non-luminous form, e.g.\ as a cold dark matter and/or a hot one. 
However, the volume of the ``missing mass'' necessary
to explain -- together with the luminous matter -- the dynamics of
large astronomical objects, considerably exceeds the masses of so far
discovered potential components of cold as well as hot dark matter.
Nevertheless, it seems that the toy model is able to supplement these
deficiencies, solving the problem of the missing mass without
introducing any additional dark matter in excess of that whose
existence has already been proved by observation.

\section{Cosmology}

\subsection{Field equations}

In this section we will examine, how the currently accepted cosmological
models are influenced by the introduction of the extra spatial
dimension. To focus our attention we will consider exclusively the
Friedmann--Robertson--Walker model(s) \cite{Wheeler}.
We suppose that our model, i.e.\ the cosmological 
toy model, will satisfy a cosmological principle which
assumes that the Universe is spatially homogeneous and isotropic, but
only for the three ``macroscopic'' spatial dimensions. Such a toy
universe is described by the metric
\begin{eqnarray}
ds^2 = \left(1 + |\lambda| a^2\right)\!  c^2 dt^2 - \left( 1 +
|\lambda| a^2\right)^{\! -1} da^2 \nonumber\\ - \!\left[
R(t)\right]^{2} \!\left[
\left( 1 - k r^2 \right)^{\! -1} d r^2
+ r^2 d \theta^2 + r^2 \sin^2 \theta \, d \varphi^2 \right] \ ,
\label{metryka}
\end{eqnarray}
where $r\in [0, \infty)$, $\theta\in [0, \pi]$, and $\varphi\in [0,
2\pi)$ are the dimensionless polar 
``co-moving'' coordinates, the curvature
parameter $k$ is a constant equal to $0$ or to $\pm 1$, and $R(t)$ 
denotes the cosmic scale factor, or the expansion parameter.

In order to solve equation (\ref{1}) with the above metric and thus to
determine conditions which should be satisfied by the scale factor
$R(t)$, the stress--energy tensor $\widetilde{T}_{\mu\nu}$ for the
investigated problem must be appropriately defined.  We assume that
\begin{equation}
{\widetilde{T}}_{\mu\nu} = T_{\mu\nu} + {\widehat{T}}_{\mu\nu} \ ,
\label{tens}
\end{equation}
where
\begin{eqnarray}
{\widehat{T}}_{\mu\nu} &=& \left( \varrho +\frac{{\widetilde{p}}}{c^2}
\right) \! u_\mu u_\nu - {\widetilde{p}} \, g_{\mu\nu} 
\;\;\;\;\;\; {\rm for} \;\; \mu = 0 , 1  \;\;\; 
\& \;\;\; \nu = 0 , 1
\label{ttens1} \\
{\widehat{T}}_{\mu\nu} &=& \left( \varrho +\frac{p}{c^2}
\right) \! u_\mu u_\nu - p \, g_{\mu\nu} 
\;\;\;\;\;\; {\mbox{otherwise,}}
\label{ttens2}
\end{eqnarray}
and $T_{\mu\nu}$ is defined by expression (\ref{t}), while
$\varrho$ denotes the rest-frame density of matter/energy of the fluid
filling the Universe, $p$ is the fluid (rest-frame) isotropic pressure
along the three ``macroscopic'' spatial dimensions, and
${\widetilde{p}}$ denotes the pressure of matter (or energy) along the
extra spatial dimension $a$, i.e.\ the pressure acting orthogonally
(on)to hypersurfaces of constant values of the coordinate $a$. In turn,
$u^\mu = d x^\mu/d \tau$ is the fluid proper ``five-velocity'' (along,
or with respect to, the $\mu$-th coordinate; $\tau$ denotes the proper
time), where $x^\mu (\tau)$ is the world-line of a fluid element. Note
that, in general, the pressures $p$ and ${\widetilde{p}}$ should not be
equal to one another as we assume spatial homogeneity and isotropy
of the Universe for the three ``macroscopic'' spatial dimensions only.

Inserting the metric (\ref{metryka}) into the Einstein equation (\ref{1})
with the stress--energy tensor given by expression ({\ref{tens}), one 
obtains the field equations which must be satisfied by the cosmic scale
factor $R(t)$,
\begin{eqnarray}
c^2 k A + {\dot{R}}^2 &=& - \frac{\kappa}{3} c^2 {\widehat{T}}_{00} R^2
\label{r1r} \\
c^2 k A + {\ddot{R}} R + {\dot{R}}^2 &=& \frac{\kappa}{3} c^2
{\widehat{T}}_{11} A^2 R^2
\label{r2r} \\
r^2 \!\left( c^2 k A + 2 {\ddot{R}} R + {\dot{R}}^2 \right) &=& \kappa
c^2 {\widehat{T}}_{33} A \ ,
\label{r3r} 
\end{eqnarray}
where we have introduced the following notation:\ $R \equiv R(t)$, 
$\dot{R} \equiv
d R(t) / d t$, $\ddot{R} \equiv d^2 R(t) / d t^2$, and $A \equiv 1 +
|\lambda| a^2$.  Obviously, the three spatial equations for the
``macro-dimensions'' $x^2 \equiv r$, $x^3 \equiv \theta$ and $x^4 
\equiv \varphi$ are equivalent to each other, 
so only one of them -- namely (\ref{r3r}) -- is written explicitely
here. In an (interesting to us) temporarily co-moving coordinate system,
for which $a= {\it const}$ and $x^i = {\it const}$, where $i=2, 3, 4$,
we obtain
\begin{eqnarray}
c^2 k A + {\dot{R}}^2 &=& - \frac{\kappa}{3} c^4 \varrho A R^2
\label{rr1} \\
c^2 k A + {\ddot{R}} R + {\dot{R}}^2 &=& \frac{\kappa}{3} c^2
\widetilde{p} A R^2
\label{rr2} \\
c^2 k A + 2 {\ddot{R}} R + {\dot{R}}^2 &=& \kappa c^2 p A R^2 \ .
\label{rr3} 
\end{eqnarray}
Since we assume that the quantity $R$ (i.e.\ the scale factor) is
exclusively a function of the coordinate time $t$ and not of the 
dimension $a$, we must assume that $k = 0$ over the whole history of 
the Universe, i.e.\ for times $t \ge 0$. 
This is because an assumption that $k = \pm 1$
inevitably results in a dependence of the factor $R$ on the dimension
$a$, or one has $\varrho (a\to \pm \infty) \neq 0$, which we would
like to avoid.  Moreover, if the scale factor $R$ is to be a function
of time $t$ only and the condition $k = 0$ is assumed, then also the
quantities $\varrho A$, $\widetilde{p} A$ and $p A$, present in
Eqs.~(\ref{rr1})--(\ref{rr3}), should not depend on $A$. Thus,
expressions determining the matter (or energy) density $\varrho$ as
well as the pressures $\widetilde{p}$ and $p$ of the fluid filling the
Universe in a function of the extra spatial dimension $a$ should take
the following form,
\begin{eqnarray}
\varrho (t=  {\it const}, a) &\sim& A^{-1} 
\label{z1} \\
\widetilde{p} (t=  {\it const}, a) &\sim& A^{-1}
\label{z2} \\
p (t= {\it const}, a) &\sim& A^{-1} \ .
\label{z3} 
\end{eqnarray}
It is clear that the above relationships reflect geometric properties
of the toy-model spacetime which is hyperbolically curved along the
additional spatial dimension $a$.(\footnote{Note that the particles of
an ideal fluid which possess a very small energy $E$ can penetrate
relatively deep into the dimension $a$ [see the formula (\ref{d_3})], 
so expressions (\ref{z1})--(\ref{z3}) do not seem to be unrealistic.}) 
In effect, we have
\begin{eqnarray} 
\varrho (t, a=0) &=& A \, \varrho (t, a) 
\label{eqeq} \\
{\widetilde{p}} (t, a=0) &=& A \, \widetilde{p} (t, a) \\ p (t, a=0)
&=& A \, p (t, a) \ .
\end{eqnarray} 

It can be easily proved that if the system of equations
(\ref{rr1})--(\ref{rr3}) is to be self-consistent for any values of $R$
and $A$, then the following condition must be satisfied,
\begin{equation}
3 p - 2 \widetilde{p} = c^2 \varrho \ .
\label{cond}
\end{equation}
Let us suppose that the fluid filling the present Universe is basically
composed of matter and that this matter exists in the form of dust,
which means that its pressure $p$ is negligible as compared with the
value of $c^2 \varrho$, i.e.\ one has $c^2 \varrho \gg p \approx 0$.
As a result we obtain the relationship $\varrho = - 2 c^{-2}
\widetilde{p}$. Obviously, this
relationship -- meaning negative value of the pressure $\widetilde{p}$
-- is an effect of the spacetime geometry of the proposed toy model
rather than of properties of the matter alone; the dependence of the
quantities $\varrho$, $p$ and $\widetilde{p}$ on the geometry of
the spacetime is especially explicitely seen in expressions
(\ref{z1})--(\ref{z3}).  Furthermore, let us note that for
radiation satisfying relationship $p = c^2 \varrho/3$ one has
$\widetilde{p} = 0$.

It is also easy to show that in the weak-field approximation, under the
condition that $c^2 \varrho \gg p \approx 0$ and for $A \approx 1$, the
$(0, 0)$-component of the Einstein equation (\ref{1}) with the
stress--energy tensor defined by expression (\ref{tens}) approaches --
while taking into account the condition (\ref{cond}) -- the Poisson
equation, $\nabla_{\bf r}^2 \psi \approx 4 \pi G
\varrho$; see also section 2.

Now we consider the (local) energy conservation, or continuity equation
for the fluid characterized by the stress--energy tensor
${\widetilde{T}}_{\mu\nu}$; this equation takes the same form as in the
case of the standard Friedmann model,
\begin{equation}
\frac{d \varrho}{d t} 
+ \left(\varrho + \frac{p}{c^2}
\right) \! \frac{3 \dot{R}}{R} = 0 \ .
\label{rc}
\end{equation}
We assume that the fluid which fills the present Universe is mainly
composed of matter existing in the form of dust. Thus we obtain
\begin{equation}
\frac{d \varrho}{d t} + \varrho
\frac{3 \dot{R}}{R} = 0 \ ,
\label{rcmd}
\end{equation}
which gives
\begin{equation}
\varrho (t, a=  {\it const}) R^3 = {\it const} \ .
\label{rhor}
\end{equation}
For radiation, which fulfils the equation of state 
$p = c^2 \varrho/3$, we have
\begin{equation}
\varrho (t, a=  {\it const}) R^4 = {\it const} \ .
\label{rhorrad}
\end{equation}

Let us define the Hubble constant $H(t) \equiv \dot{R} (t) / R(t)$, the
critical density of matter
\begin{equation}
\varrho_c \equiv - \frac{3 H^2}{\kappa c^4 A} \ ,
\end{equation}
and the current critical density of matter
\begin{equation}
\varrho_{c \, 0} \equiv \varrho_c (t_0) 
= - \frac{3 H_0^2}{\kappa c^4 A} =
\frac{3 H_0^2}{8 \pi G A}
\end{equation}
where $H_0 \equiv H(t_0)$ denotes the present value of the Hubble
constant.  Let $\varrho_0$ represents the density of matter in the
present Universe, so $\varrho_0 \equiv \varrho (t_0)$. 

We can also define a parameter $\Omega$, frequently used in 
cosmology, as
\begin{equation}
\Omega \equiv \frac{\varrho}{\varrho_c} \ ,
\end{equation}
so we have
\begin{equation}
\Omega = - \frac{\kappa c^4}{3 H^2} \varrho A \ .
\label{omega}
\end{equation}
Furthermore, the present value of the density parameter $\Omega$ reads
\begin{equation}
\Omega_0 \equiv \Omega(t_0) = \frac{\varrho_0}{\varrho_{c \, 0}} \ .
\label{om0}
\end{equation}

It can be easily noticed that for a given density of matter/energy
$\varrho$, the parameter $\Omega$ varies by a factor of $A$ with 
respect to its definition in the standard model, since we have
\begin{equation}
\Omega^{\it stand} = - \frac{\kappa c^4}{3 H^2} \varrho \ ,
\label{olddef}
\end{equation}
and then one obtains
\begin{equation}
\Omega = A \, \Omega^{\it stand} \ .
\label{omom}
\end{equation} 
It is of course an effect of the modification of the critical density
value with respect to its definition in the original Friedmann model,
\begin{equation}
\varrho_c = A^{-1} \varrho^{\it stand}_c \ .
\end{equation}

Let us remind the reader that the assumption $R =R(t)$ -- so $R$ is 
independent of the coordinates other than $t$ 
-- leads to the conclusion that the value of
$k$ is equal to $0$. Therefore -- from Eq.~(\ref{rr1}) -- one can
obtain an important relationship,
\begin{equation}
\Omega = 1 \ .
\label{om1}
\end{equation}
Hence, the density of matter/energy in the Universe assumes its
critical value, meaning that the geometry of the three ``macroscopic''
spatial dimensions of the Universe is flat. Obviously,  
the flat geometry of these ``macroscopic''
spatial dimensions is a consequence of 
the assumption that $k = 0$, which {\it de facto} 
is equivalent to the formula (\ref{om1}).

Note that according to the equalities (\ref{omom}) and (\ref{om1}), 
the following can be written,
\begin{equation}
\Omega^{\it stand} = A^{-1} \ .
\label{omst}
\end{equation} 
Then a question arises as to whether -- bearing 
in mind relations (\ref{om1})
and (\ref{omst}) -- we can estimate the value of the cosmological
density parameter defined as in the standard model, i.e.\ $\Omega^{\it
stand}$.  It seems to be difficult as we got used to the problems with
the estimation of the value of the factor $A$.  At this place we should
appeal to the experiment\footnote{Note that in the toy model the
quantity $\Omega$ only is of an essential theoretical significance,
whereas $\Omega^{\it stand}$ is the quantity of a rather purely
experimental relevance. As one may see, there exist two approaches to
estimating the ``actual'' value of the quantity 
$\Omega^{\it stand}$:\ the first one -- experimental -- on the basis 
of the formula (\ref{olddef}), and the second one -- ``theoretical'' 
-- with the use of expression (\ref{omst}). Both the approaches
should in principle give similar results. In practice, however, they
can differ depending e.g.\ on which is the method used to estimate the
density of matter $\varrho$; see section 5.2.} and observe that while
experimentally estimating the value of the quantity $\Omega^{\it stand}$ 
applying the formula (\ref{olddef}), the density of matter $\varrho$ is
usually calculated as, roughly speaking, the quotient of the detected
mass -- i.e.\ of the mass which is subject to our perception -- by the
(three-dimensional) volume of the space with $a = 0$, within which
(i.e.\ for $a = 0$) as well as around which (i.e.\ for $a \neq 0$) the
detected mass is situated.  The value of the factor $A$ in the formula
(\ref{omst}) should then correspond to the range, or width of the
additional spatial dimension $a$ (around $a=0$, on both sides of $a$,
i.e.\ for $\pm a$) which is observable, or rather detectable indirectly,
i.e.\ by detecting the mass which is situated there.  Let us note that
$\Omega \ge \Omega^{\it stand}$ always, since $A \ge 1$, as one has
$|\lambda| a^2 \ge 0$ . We expect that for the present Universe
as a whole, the quantity $A$ could be of the order of $10^{1}$, or even
of $10^{2}$. Namely, for such values of $A$ the density of matter
$\varrho (A=10)$ is, according to the formula (\ref{eqeq}), an order of
magnitude lower than the density of matter for $A = 1$, i.e.\ for $a =
0$. Thus, the visibility of the matter situated at $A = 10$ -- or the
perception of it -- is much weaker than that of the matter concentrated
at $A = 1$. In other words, one has $\varrho (t, A_d = 10) = 10^{-1}
\varrho (t, A = 1)$, so most
matter -- with respect to the dimension $a$ -- is situated in the
region of space around $a=0$, within the interval $a \in [ - a_d,
a_d]$, where $A_d \equiv 1 + |\lambda| a^2_d$.  Thus, according to
the formula (\ref{omst}), it could be assumed that
\begin{equation}
\Omega^{\it stand} \sim A_d^{-1} = 10^{-1} \ .
\end{equation}
One may suppose that the given above value of $A_d$ -- one order of
magnitude greater than $A(a=0)$ -- should roughly correspond to the
range of values of the additional spatial coordinate, given by 
$a \in [ - a_d, a_d]$, which is subject to our direct perception.

Another, more direct and probably more appropriate approach to
estimating the value of the quantity $\Omega^{\it stand}$ could proceed
as follows: Let us assume that the quantities $\pm a_d$ correspond to
the limits of the directly observable interval of the dimension 
$a$ around $a = 0$, so one has $\Omega^{\it
stand} = A_d^{-1} = (1 + |\lambda| a_d^2)^{-1}$.  If we reasonably
assume that $a_d = L = 2 \pi/|\lambda|^{1/2}$, then we obtain
\begin{equation}
\Omega^{\it stand} = \left( 1 + 4 \pi^2 \right)^{\! -1} \cong 0.025 \ .
\label{omstand}
\end{equation}

One more approach to estimating the value of the parameter 
$\Omega^{\it stand}$ -- the most thorough and perhaps the most 
promising -- could be
imagined: Let us assume that all the observable matter is
concentrated within the interval of the additional spatial dimension
given by $a \in [ - a_d, a_d]$ or, in other words, within the
interval $A \in [ 1, A_d]$ ``taken twice'', as one has $A_d = 1 +
|\lambda| (\pm a_d)^2$. We may then write 
\begin{equation}
\Omega^{\it stand} \propto 2 \int_1^{A_d} \varrho_a \, d A
= 2 \varrho_{a=0} \int_1^{A_d} A^{-1} \, d A = 2 \varrho_{a=0}
\ln A_d 
\label{m3a} 
\end{equation}
where $\varrho_a \equiv \varrho (a)$. In the case of the above 
expression, for the normalization factor one can take
\begin{equation}
\Omega \propto 2 \int_1^{A_d} \varrho_a A \, d A = 2 \varrho_{a=0}
\int_1^{A_d} dA = 2 \varrho_{a=0} \left(A_d - 1\right) \ ; 
\label{m3b}
\end{equation} 
note that, while obtaining both the above formulae, we have applied
the equality (\ref{eqeq}) which results, for instance, in the 
relationship $A d \varrho_a = - \varrho_a d A$ for $t = {\it const}$.
Dividing expressions (\ref{m3a}) and (\ref{m3b}) by one
another and taking into account the formula (\ref{om1}), we find that
\begin{equation}
\Omega^{\it stand} = \frac{\ln A_d}{A_d - 1} \ .
\label{ad_2}
\end{equation}
If we then assume that $a_d = L$, so $A_d = 1 + 4 \pi^2$, 
we obtain
\begin{equation}
\Omega^{\it stand} = 
\frac{\ln \!\left( 1 + 4 \pi^2 \right)}{4\pi^2} \cong 0.094 \ .
\label{ad_3}
\end{equation}
However, if we demand the value of $A_d$ to fulfil the condition
$\varrho (t, A_d) = 10^{-1} \varrho (t, A = 1)$, then we obtain
that $A_d = 10$, and from the formula (\ref{ad_2}) we have
\begin{equation}
\Omega^{\it stand} \cong 0.256 \ ,
\label{omdm}
\end{equation}
which remains in agreement with the commonly accepted value
of the cosmological mass density parameter $\Omega_0^{\it stand}$
that incorporates a contribution usually attributed to the existence 
of dark matter; see, for instance, Ref.~\cite{Turner} and references 
therein. In the toy model, such a contribution is a consequence of the 
existence of the additional spatial dimension $a$, and this has been
explicitely taken into account while calculating the above value
of the cosmological density parameter $\Omega^{\it stand}$; see also 
sections 4.2 and 5.2. Thus, the above considerations indicate, to some 
extent, the self-consistency of the model proposed in this paper.

However, we can also calculate the cosmological density parameter
$\Omega^{\it stand}$ in a quite similar manner as above, but with
the appropriate integrations performed now over the quantity $d a$ 
instead of $d A$, 
\begin{eqnarray}
\Omega^{\it stand} & \propto & 
\int_{- \sqrt{|\lambda|} \, a_d}^{+ \sqrt{|\lambda|} \, a_d} 
\varrho_a \, d \!\left( \sqrt{|\lambda|} \, a \right)
= 2 \varrho_{a=0} \, {\rm arc} \tan \!\left( \sqrt{|\lambda|} \, a_d 
\right) \label{m4a} \\
\Omega & \propto & 
\int_{- \sqrt{|\lambda|} \, a_d}^{+ \sqrt{|\lambda|} \, a_d}
\varrho_a A \, d \!\left( \sqrt{|\lambda|} \, a \right)
= 2 \varrho_{a=0} \sqrt{|\lambda|} \, a_d \ ,
\label{m4b}
\end{eqnarray} 
so after the normalization of the parameter $\Omega$ to unity one 
arrives at the formula
\begin{equation}
\Omega^{\it stand} = \frac{\, {\rm arc} \tan \!\left( \sqrt{|\lambda|} 
\, a_d \right)}{\sqrt{|\lambda|} \, a_d} \ .
\label{ad_4}
\end{equation}
Assuming that $a_d = L$, so $|\lambda|^{1/2} a_d = 2 \pi$, we
obtain $\Omega^{\it stand} \cong 0.225$, whereas for $A_d = 10$,
that is when $|\lambda|^{1/2} a_d = 3$, one has $\Omega^{\it stand}
\cong 0.416$. It is then clear that the whole above discussion 
concerning the estimation of the actual value of the cosmological 
density paramater $\Omega^{\it stand}$ seems to confirm the fact
that this quantity is not of essential relevance in the toy model. 

\bigskip
\smallskip

Now let us define a quantity called the deceleration parameter,
\begin{equation}
q \equiv - \frac{{\ddot{R}} R}{{\dot{R}}^2} \ .
\end{equation}
According to Eqs.~(\ref{rr1}) and (\ref{rr3}), in the case of dust
filling the Universe for which $p = 0$ and $\widetilde{p} = - c^2
\varrho/2$, 
there holds the relationship $q = \Omega/2$. In turn, one finds $q =
\Omega$ for radiation.

It should also be noted that the toy model does not alter the formula
determining the present age of the Universe, which remains of a
familiar form,
\begin{equation}
t_0 = \frac{1}{H_0} \int^1_0 \frac{dx}{\left( 1 - \Omega_0 +
\Omega_0 x^{-1} \right)^{1/2}} \ .
\label{wiek}
\end{equation}
[The suffix ``$0$'' in expression (\ref{wiek}) and in the rest of the 
paper denotes the present value of any given quantity.] Since in the toy
model $\Omega_0 =1$, then the present age of the Universe is equal to
$t_0 = 2/(3 H_0)$.  The latest research estimates the Hubble constant
value to be $H_0 = 65 \pm 10$ km s$^{-1}$ Mpc$^{-1}$; see
Ref.~\cite{Freedman}. Since the ages of the oldest globular clusters
are estimated approximately to amount to $11.5 \pm 1.3$ Gyr
\cite{Chaboyer1}, 
then the age of the Universe obtained from the formula (\ref{wiek}) 
with $\Omega_0 =1$ is not inconsistent with the observational data;
see also Ref.~\cite{Chaboyer2}.

\subsection{Cosmological density parameters}

This section contains considerations concerning
defined in cosmology so-called
total density parameter which is given as a sum of the mass density
parameter $\Omega$ and a ``vacuum'' density parameter
$\Omega_{\Lambda}$, the latter associated with the possible existence
of a ``macroscopic'' cosmological constant $\Lambda$; we define the
current value of the total cosmological density parameter as
\begin{equation}
\Omega_{\it total} \equiv \Omega_0 + \Omega_{\Lambda}
\label{total}
\end{equation}
where
\begin{equation}
\Omega_\Lambda \equiv \frac{\Lambda c^2}{3 H_0^2} A =
A \, \Omega^{\it stand}_\Lambda \ .
\end{equation}
The value of the parameter $\Omega_\Lambda$ has a significant impact on
other cosmological quantities, like for instance the age of the
Universe $t$ or the deceleration parameter $q$; note that in
the considerations presented in section 5.1 we have assumed that
$\Omega_\Lambda =0$.

\subsubsection{Cosmological mass density parameter $\Omega_0$}

First let us note that the vast majority of experimental cosmology
measurements is related to the cosmological density parameter defined
as in the standard model, i.e.\ to the quantity
$\Omega^{\it stand}_0$.  In order to
determine its value it is not necessary -- at least from the
``experimental'' point of view -- to know the value of the factor $A$;
see the formula (\ref{olddef}).\footnote{Note that while experimentally
estimating the value of the quantity $\Omega^{\it stand}_0$ applying
the formula (\ref{olddef}), the density of matter $\varrho_0$ is usually
calculated as, roughly speaking, the quotient of the directly and/or
indirectly detected mass by the (three-dimensional) volume of the space
with $a = 0$, within which (i.e.\ for $a = 0$) as well as around which
(i.e.\ for $a \neq 0$) the detected mass is situated.} Considerations
presented in this subsection principally concern the parameter
$\Omega_0^{\it stand}$, which depicts the density of matter occurring
in the contemporary Universe, but neglects at the same time the
existence of the extra spatial dimension, by omitting the factor $A$ in
the definition of the quantity $\Omega_0$; see expressions 
(\ref{omega})--(\ref{omom}).

On the grounds of the analysis of a number of observational data one
might say that the cosmological density parameter $\Omega_0^{\it
stand}$ takes the value from the interval $(0.01, 1.1)$. 
The value of the quantity
$\Omega_0^{\it stand}$ was estimated on the level $0.2 - 0.3$ based,
among others, on the extensive studies of the dynamics of large
astronomical objects, like clusters of galaxies; see, for instance,
Refs.~\cite{Carlberg,Bahcall,Bahcall1,Donahue} and references therein.
One should ask, however, whether the value of the cosmological density
parameter $\Omega_0^{\it stand} \in (0.2 , 0.3)$ determined in this 
manner is true, i.e.\ is it the value reflecting the real density of 
matter in the present Universe.

Let us note that in order to determine the value of the parameter
$\Omega_0^{\it stand}$ by means of the analysis of the dynamics of
large astronomical objects like mentioned clusters of galaxies, first
of all, the mass(es) of investigated object(s) should be estimated.
This, in turn, involves an appeal to the virial theorem, which has
been described briefly in section 4.2 of this paper. However, for the
determination of the mass of investigated object(s) -- see
Refs.~\cite{Carlberg,Bahcall,Bahcall1,Donahue} and references therein
-- equation (\ref{dm1}) of the standard Newtonian theory is used
instead of equation (\ref{dm2}) which results from the toy model.
Consequently, the value of the object's mass so obtained is greater than
that which would be determined within the framework of the toy model,
i.e.\ applying Eq.~(\ref{dm2}).  Thereby, the value of the cosmological
density parameter $\Omega_0^{\it stand}$ obtained on the grounds of the
discussed toy model can be considerably smaller than the value of
$\Omega_0^{\it stand}$ determined by means of the standard theory which
does not take into account the presence of the extra spatial dimension.
Let us note, however, that the accurate determination of the value of
the parameter $\Omega_0^{\it stand}$ based on the application of the
virial theorem in terms of the toy model, seems to be very difficult in
view of the fact that we do not know the value of the energy $U$ of the
investigated object.

Hence, as far as we assume the possibility that the dark matter does
not exist in the Universe in any significant amount, the real value of
the parameter $\Omega_0^{\it stand}$ may be considered to be close to
the value implied by the primordial baryon nucleosynthesis theory,
which value is estimated to be $\Omega_B^{\it stand} \approx 0.020 \pm
0.007 h^{-2}$, where $h = H_0/(100$ km s$^{-1}$ Mpc$^{-1}$); see
Ref.~\cite{Hogan} and references therein. This value is in accord with
the recent estimations of the amount of the baryon matter in the
present Universe, giving $0.007 \le \Omega_B^{\it stand}
\le 0.041$ with a central value $\Omega_B^{\it stand} \approx 0.021$ 
\cite{Hogan}; compare this result with the formula (\ref{omstand}). 
Note that, in the case when the value of the parameter $\Omega_0^{\it
stand}$ is so small that one has $\Omega_0^{\it stand} \approx
\Omega_B^{\it stand}$, 
then the value of the deceleration parameter $q_0^{\it stand} =
\Omega_0^{\it stand}/2$ is very close to zero 
(i.e.\ of the order of $10^{-2}$).  Such a small value of the
deceleration parameter $q_0^{\it stand}$ would remain consistent with
recently obtained experimental data \cite{Rines}.

We should also note that estimating the value of the cosmological
density parameter $\Omega_0^{\it stand}$ to be of the order of $10^{-1}
- 10^{-2}$ allows one, according to the formula (\ref{omst}), to 
estimate the quantity $A$ for the Universe to be of the order of 
$10^{1} - 10^{2}$, which
agrees with the order of the quantity $A$ which we assumed in section
5.1 of this paper. Thus, both the approaches:\ the theoretical one,
investigated in section 5.1, and the experimental one, described in 
the present subsection, seem to remain consistent, 
as they give similar results.

It has already been mentioned that the determination of the real value
of the cosmological density parameter $\Omega_0^{\it stand}$ in terms
of the toy model seems to be very complicated. This is because of the
fact that in many cases it depends on the knowledge of an {\it a
priori} unknown value of the energy $U$, which may in principle vary
for different objects. It looks, however, that the value of the mass of
the investigated object(s) and then the value of the parameter
$\Omega_0^{\it stand}$ can be determined in a relatively accurate manner
by the extensive investigation of gravitational microlensing events,
concerning both a luminous and a non-luminous matter. From section 4.1 
of this paper we already know that there does not exist any additional
light deflection caused by the existence of the extra spatial
dimension.  Therefore, in this case, the lack of knowledge of the
parameter $U$ does not really matter.

And indeed, recently there have been undertaken numerous attempts at
determining the mass of large astronomical objects that were
based on the gravitational lensing effect; see
Refs.~\cite{Smail,Luppino,Bahcall2,Hjorth}.  In several works, however,
e.g.\ in Refs.~\cite{Bahcall2,Hjorth}, a final determination of the
cosmological density parameter $\Omega_0^{\it stand}$ involved the use
of models assuming the existence of a substantial amount of dark
matter,\footnote{The models which assume the existence of a substantial
amount of dark matter lead, among others, to the following relation
$\sigma_8 (\Omega_0^{\it stand})^{0.5} \simeq 0.5$, linking together
the value of the mass density parameter $\Omega_0^{\it stand}$ and the
value of the amplitude of mass fluctuations, $\sigma_8$; see
Refs.~\cite{Eke,Viana1,Pen,Kitayama}.} e.g.\ by applying the virial
theorem in the form of Eq.~(\ref{dm1}) rather than of Eq.~(\ref{dm2}).
It seems that referring to the models
assuming the existence of large amounts of dark matter in the Universe
considerably limits the possibility of a completely objective -- 
i.e.\ not depending on the applied method -- determination of the 
value of the cosmological density parameter $\Omega_0^{\it stand}$.

\bigskip

Summarizing, it seems that the measurement of the mass of investigated
object(s), which is necessary for the determination of the quantity
$\Omega_0^{\it stand}$, should not be performed by means of the
analysis of the investigated object's dynamics. The mass should rather 
be obtained, for example, by studying gravitational microlensing events, 
or from the analysis of the luminosity and mass-to-light ratio for
investigated object(s).  Among other interesting methods of the mass
determination, used for instance to find masses of clusters of
galaxies, is the analysis of the galaxy cluster X-ray integral 
temperature distribution function \cite{Viana1,Viana2}.

Thus, in the context of this subsection, it seems to be reasonable to
theoretically estimate the value of the quantity $\Omega_0^{\it stand}$
in such a manner that it incorporates the mass of the whole 
{\it observable} matter only, i.e.\ the mass of that matter 
which is concentrated within the range
of the additional spatial dimension $a$, being subject to the direct
perception of us, or rather of our measuring equipment; see
the formula (\ref{omstand}) and/or expressions
(\ref{ad_2}) and (\ref{ad_3}) in section 5.1.

\subsubsection{Cosmological ``vacuum'' density parameter $\Omega_\Lambda$}

In this section we will discuss estimations regarding the values of the
cosmological density parameters, obtained on the basis of observations
of luminous sources known as type Ia supernovae; see
Refs.~\cite{Riess,Perlmutter} and references therein.  
The analysis of the observational data
concerning these objects leads to the conclusion that the values of
both the cosmological parameters $\Omega_0$ and $\Omega_\Lambda$ are
considerably greater than zero. On the basis of a number of 
measurements there were found, among others, the following values 
of the cosmological density parameters:
\begin{equation}
(\Omega_0 , \Omega_\Lambda) \approx (0.24 , 0.72), (0.00 , 0.48), 
(0.80 , 1.56), (0.72 , 1.48);
\label{c1c}
\end{equation}
see Ref.~\cite{Riess}.  All of the above values were obtained by means 
of two different light-fitting methods:\ the first two by means of the
so-called MLCS method, whereas the next two -- with the use of the
template fitting approach. In turn, in the recent paper of
Perlmutter {\it et al.} \cite{Perlmutter}, concerning the analysis
of data on $42$ high-redshift type Ia supernovae, there has been
estimated a relationship between the cosmological parameters $\Omega_0$
and $\Omega_\Lambda$, which has the form
\begin{equation}
0.8 \, \Omega_0 - 0.6 \, \Omega_\Lambda \approx - 0.2 \pm 0.1 \ .
\label{c1}
\end{equation}

A question arises here as to whether any of the above estimations
approaches the real values of the cosmological density parameters. 
In an attempt to answer it, first let us note that the theoretical
analysis of the observational data presented in
Refs.~\cite{Riess,Perlmutter} is based on the so-called apparent
magnitude--redshift relation; see Refs.~\cite{Goobar,P97} and 
references therein.

Let $M$ and $m$ denote the absolute and apparent magnitude, 
respectively, at a given redshift $z$ for a luminous object. 
Results of measurements of both the
quantities $M$ and $m$ depend on values of the cosmological density
parameters $\Omega_0$ and $\Omega_\Lambda$ in a way described by the
apparent magnitude--redshift relation which reads
\begin{equation}
m (z) = M - 5 \log H_0 + 25 + 5 \log \left[ {\cal D}_L \!\left( z ;
\Omega_0 , \Omega_\Lambda\right)\right] + K \ ,
\label{mrr}
\end{equation}
where $\log x \equiv \log_{10} x$. The quantity $K$ in the above
formula denotes the so-called $K$-correction which appears in the above
equation because the emitted and the detected photons coming from the
receding object have different wavelengths; see
Refs.~\cite{Kim2,Kim1} and references therein.  The quantity ${\cal
D}_L$, where ${\cal D}_L \equiv d_L H_0$ and $d_L$ denotes the
luminosity distance \cite{Peebles}, is a function of the variables
$z$, $\Omega_0$ and $\Omega_\Lambda$
\cite{Goobar}; it should be emphasized that the function ${\cal D}_L$ does
not depend on the value of the Hubble constant $H_0$. It can be easily
shown that in the case of the toy model, under the assumption that
$\Omega_0 =1$ and $\Omega_\Lambda =0$, this function takes the
following form,
\begin{equation}
{\cal D}_L (z; \Omega_0 = 1, \Omega_\Lambda = 0) = c\, (1+z)
\int^z_0 \frac{dz'}{\left( 1 + z' \right)^{3/2}} \ ,
\label{dtoy}
\end{equation}
whereas the general expression for ${\cal D}_L$ in the Friedmann
standard model reads
\begin{eqnarray}
{\cal D}_L (z;\Omega_0,\Omega_\Lambda) &=& \frac{c\,
(1+z)}{|\Omega_k|^{1/2}} \sin_k
\!\left( |\Omega_k|^{1/2} \right.
\nonumber\\ &\times& \left.
\int^z_0 \frac{dz'}{\left[
\Omega_0 \!\left( 1 + z' \right)^{3} + \Omega_k 
\!\left( 1 + z' \right)^2 +
\Omega_\Lambda\right]^{1/2}}\right) \ ,
\label{dstand}
\end{eqnarray}
where $\Omega_k \equiv 1 - \Omega_0 - \Omega_\Lambda$ and $\sin_k x$
is equal to $\sinh x$ if $\Omega_k > 0$, to $\sin x$ if $\Omega_k <
0$, or to $x$ if $\Omega_k = 0$. Note that, while deriving the formula
(\ref{dtoy}), we have put $A = 1$ for radial light ray(s)
connecting the investigated supernova(e) and the observer. This is
because, in the considered problem, the behaviour of the light 
particles' trajectories with respect to the dimension $a$
is not important, but only in the spacetime $a = 0$; thus, the 
assumption that $A = 1$ seems to be relevant here.

Moreover, there is usually defined a quantity called the ``Hubble
intercept'',
\begin{equation}
{\cal M} \equiv M - 5 \log H_0 + 25 \ .
\label{hi}
\end{equation}
To some extent, it can be measured directly, i.e.\ without knowing
$H_0$, if we only know the apparent magnitude $m$ for objects showing
relatively low redshifts $z$. In such a case, the relationship
(\ref{mrr}) takes the form
\begin{equation}
m (z) = {\cal M} + 5 \log (c z) + K \ .
\label{red}
\end{equation}
Thus, the quantity $\cal M$ can be obtained from low-redshift apparent
magnitude measurements, i.e.\ from measurements of the apparent
magnitudes and redshifts of low-redshift objects. Performing
low-redshift measurements for objects of a similar type (e.g.\ type
Ia supernovae) for the purpose of determining $\cal M$ and then 
carrying out measurements of the apparent 
magnitudes $m$ and redshifts $z$ for a
number of distant, high-redshift objects of the same type, we can
obtain the best-fit values of the cosmological density parameters
$\Omega_0$ and $\Omega_\Lambda$ to solve Eq.~(\ref{mrr}); see
Ref.~\cite{Goobar}.

It is known, however, that there exist perturbations on a spatially
homogeneous and isotropic Universe.  These perturbations can cause
the value of the Hubble constant $H_0^L$ measured locally (redshift
$z\le 0.05$) to be different from the value of the global ($z > 0.3$)
Hubble constant $H_0^G$. Note that in such a case, a definition of the
directly measurable quantity $\cal M$ should take the following form,
\begin{equation}
{\cal M} \equiv M - 5 \log H_0^L + 25 \ .
\label{c2}
\end{equation}
Then, in the case when $H_0^L \ne H_0^G$, for the analysis of data
concerning high-redshift type Ia supernovae the following formula
should be applied,
\begin{eqnarray}
m (z) &=& M - 5 \log H_0^G + 25 + 5 \log \left[ {\cal D}_L (z; \Omega_0
= 1, \Omega_\Lambda = 0)\right] + K \nonumber\\ &=& {\cal M} + 5 \log
\frac{H_0^L}{H_0^G} + 
5 \log \left[ {\cal D}_L (z; \Omega_0 = 1,
\Omega_\Lambda = 0)\right] + K \ .
\label{mrr1}
\end{eqnarray}
However, for this purpose equation (\ref{mrr}) was used in 
Refs.~\cite{Riess,Perlmutter}, which equation may
be written in the following form,
\begin{equation}
m (z) = {\cal M} + 5 \log \left[ {\cal D}_L \!\left( z ;
\Omega_0^{\it app} , \Omega_\Lambda^{\it app}\right)\right] + K \ ;
\label{app}
\end{equation}
[the superscript {\it app} comes from the author of this paper.] 
It seems then that the quantities $\Omega_0^{\it app}$ and 
$\Omega_\Lambda^{\it app}$ are not the real cosmological parameters 
$\Omega_0$ and $\Omega_\Lambda$.  If we compare Eq.~(\ref{mrr1}) with 
Eq.~(\ref{app}), then it can be easily noticed that the quantities 
$\Omega_0^{\it app}$ and $\Omega_\Lambda^{\it app}$ depend in an 
essential way on the value of the quotient $H_0^L/H_0^G$, and also 
on the value of redshift $z$.

Deriving a formula for ${\cal D}_L$ in the case of the toy model, we
have assumed that $\Omega_0 = 1$ and $\Omega_\Lambda = 0$; see the 
equality (\ref{dtoy}).  For the values of the apparent cosmological 
density parameters $(\Omega_0^{\it app} , \Omega_\Lambda^{\it app})$ 
equal to $(0.24 , 0.72)$, $(0.00 , 0.48)$, $(0.80 , 1.56)$ and 
$(0.72 , 1.48)$ \cite{Riess} we obtain 
-- from the comparison of Eq.~(\ref{mrr1}) with Eq.~(\ref{app}), with 
taking into account the formulae (\ref{dtoy}) and (\ref{dstand}) -- the
following values of the quotient $H_0^L/H_0^G$:\ 1.219, 1.223, 1.270 and
1.273, respectively. For the high-redshift supernovae investigated in
Ref.~\cite{Riess} we have assumed here the average value of redshift
$z_{\it av} \approx 0.5$.  It turns out that for substantial
differences between the values of the apparent cosmological parameters
$\Omega_0^{\it app}$ and $\Omega_\Lambda^{\it app}$ obtained by means
of the two light-fitting methods, the values of the quotient
$H_0^L/H_0^G$ remain almost the same 
for each of the light-fitting methods separately. 
Therefore, one should consider the hypothesis that the
real values of the cosmological parameters $\Omega_0$ and
$\Omega_\Lambda$ are close to $1$ and $0$, respectively -- as
predicted by the toy model, whereas for objects with a small 
redshift $z$ the value of the locally measured Hubble constant $H_0^L$ 
is slightly greater than its global value $H_0^G$.  To some extent, 
this effect may occur due to the peculiar streaming motion in our 
neighbourhood which is toward the so-called Great Attractor,
situated at the redshift $z \sim 0.02$; see
Ref.~\cite{Mould} and/or Fig.~5.4 in Ref.~\cite{Peebles}.

Now let us discuss results presented in Ref.~\cite{Perlmutter}. If we
take the average redshift for the investigated objects to be equal to
$z_{\it av} \approx 0.5$, then we can determine the value of the quotient
$H_0^L/H_0^G$ for different values of the parameters $\Omega_0^{\it
app}$ and $\Omega_\Lambda^{\it app}$ which satisfy Eq.~(\ref{c1}). For
instance, for $(\Omega_0^{\it app} ,
\Omega_\Lambda^{\it app}) = (0 \pm 5/80 , 1/3 \pm 1/12)$
we obtain $H_0^L/H_0^G = 1.193 \pm 0.029$.  The value of the quotient
$H_0^L/H_0^G$ grows very slowly with increasing values of the
cosmological parameters $\Omega_0^{\it app}$ and $\Omega_\Lambda^{\it
app}$, becoming equal e.g.\ to $H_0^L/H_0^G = 1.241
\pm 0.035$ for $(\Omega_0^{\it app} , \Omega_\Lambda^{\it app}) = (1 
\pm 5/80, 5/3 \pm 1/12)$. 
Such a small difference between the values of the quotient
$H_0^L/H_0^G$ for considerably varying values of the apparent
cosmological density parameters $\Omega_0^{\it app}$ and
$\Omega_\Lambda^{\it app}$ which satisfy empirically obtained
Eq.~(\ref{c1}), seems to confirm 
the hypothesis that the real values of the
cosmological density parameters $\Omega_0$ and $\Omega_\Lambda$ are
close to unity and zero, respectively. The value of the local Hubble
constant $H_0^L$ for objects with small redshifts ($z \le 0.05$)
whereas is somewhat greater than the value of the global Hubble
constant $H_0^G$.

However, any binding verification of this hypothesis is not easy,
especially if one keeps in mind the difficulties in independently
of one another determining the four quantities:\ $H_0^L$, $H_0^G$,
$\Omega_0$ and $\Omega_\Lambda$. In the case of the analysis of the
observational data that concern type Ia supernovae, usually two of the
above quantities had been {\it a priori} fixed, which in the next step
allowed one the determination of the remaining two other parameters. 
For instance, in Refs.~\cite{Riess,Perlmutter} it had been assumed that
$H_0^L = H_0^G$ and the attempts were made to determine the values of
$\Omega_0$ and $\Omega_\Lambda$ or -- to be more precise -- of
$\Omega_0^{\it stand}$ and $\Omega_\Lambda^{\it stand}$. Alternatively,
in Refs.~\cite{Kim2,Kim1} the quantities $\Omega_0$ and $\Omega_\Lambda$ 
were left as free (changing) parameters and the authors instead
tried to determine a value of the quotient $H_0^L/H_0^G$, each time for
some fixed values of the quantities $\Omega_0$ and $\Omega_\Lambda$,
i.e.\ as a (two-variable) function of both the quantities $\Omega_0$ and
$\Omega_\Lambda$.

Moreover, let us note that -- in general -- different observed ``local
supernovae calibrators'' (i.e.\ low-redshift supernovae) may lie within
different cosmological local or sub-local flows (which is, of course,
rather improbable for a set of supernovae situated within a
sufficiently small region of space). This could mean that for different
local supernovae the quantity $H_0^L$ might have varying values.  
On the other hand, it
cannot be ruled out that also some of the observed high-{\it z}
supernovae lie within certain local cosmological flows; of course,
values of the Hubble constant for such objects might differ from the
value of the global Hubble constant $H_0^G$.  It then seems that, in
order to come to a conclusion concerning the real values of the
parameters $\Omega_0$ and $\Omega_\Lambda$, many more data on the
spatial distribution of the value of the Hubble constant are needed.

Despite all the above objections and difficulties in determining
the real values of the cosmological parameters $\Omega_0$ and
$\Omega_\Lambda$ based on the supernovae observational data, it should
be stated that the method presented in this subsection is -- if one
assumes the existence of the extra spatial dimension -- much more
promising than studying of the dynamics of large astronomical objects.
It is caused by the fact that in none of the equations and formulae
applied in this subsection does there seem to be any dependence upon the
{\it a priori} unknown quantity $U$, whose occurrence complicates
considerably attempts to determine the real values of the
cosmological density parameters in the case of analysing the dynamics
of large astronomical objects by means of the virial theorem.

\subsubsection{Fluctuations of cosmic microwave background radiation}

Assuming that $\Omega_0 =1.0$, we can estimate the value of the
cosmological density parameter $\Omega^{\it stand}_0$ defined as in the
standard model, which is connected to the value of the quantity
$\Omega_0$ by the relation (\ref{omom}). Taking into account that for
the present Universe as a whole the value of the factor $A$ can be of
the order of $10$, one can suppose 
that $\Omega^{\it stand}_0 \sim 10^{-1}$.
Could the above statement be an important fact that might help to
explain some discrepancies between various observational data as well
as between some experimental and theoretical results (the latter, for
instance, coming from a computer simulation)?  According to the latest
reports, the value of the cosmological mass density parameter close to
$0.1$ (see section 5.2.1), on the assumption that $\Omega_\Lambda = 0$
(see section 5.2.2), is not consistent with the observed fluctuations 
of the cosmic microwave background radiation (CMBR).  

The value and shape of the CMBR fluctuations, and in effect also of
primordial fluctuations in the density of matter, are reflected in the
angular power spectrum of CMBR. As it turns out, the position of the
first acoustic (Doppler) peak at the Legendre multipole in the angular
power spectrum of the temperature fluctuations of CMBR is almost
exclusively determined by the value of the cosmological density
parameter $\Omega_0$, and varies as
\begin{equation}
\ell_{\it peak} \simeq \frac{200}{\sqrt{\Omega_0}} \ ;
\label{peak}
\end{equation}
this occurs because the angular scale $\ell_{\it peak}$ of the main 
peak reflects the size of the horizon at last scattering of the CMBR
photons; see Ref.~\cite{MNRAS} and references therein as well as
Ref.~\cite{Peebles}.  It should be stressed that the quantity $\Omega_0$ 
in the formula (\ref{peak}) corresponds in the toy model to the
cosmological mass density parameter defined in expression (\ref{om0}), 
and not to the cosmological mass density parameter $\Omega_0^{\it stand}$
defined as in the standard model and given by the relation $\Omega^{\it
stand}_0 = A^{-1} \Omega_0$; roughly speaking, this occurs, since the
factor $\dot{R}/R$ -- which enters the formula for the optical depth to
the surface of the Thomson (``last'') scattering by free electrons
\cite{Peebles} -- 
is proportional to $\Omega$, and not to $\Omega^{\it stand}$; see
Eq.~(\ref{rr1}) and expressions (\ref{omega}), (\ref{olddef}).

Based on studies of the power spectrum of CMBR, it can be stated that
the observed fluctuations of CMBR are too small to have been produced
in an open Universe with small values of $\Omega_0$ like, for instance,
$\Omega_0 = 0.3$.  It turns out, that only values $\Omega_0 > 0.4$
remain consistent with the degree of fluctuations of the cosmic
microwave background radiation as well as with the recent estimations
concerning the Hubble constant value \cite{MNRAS}.

A new limit on the value of the cosmological parameter $\Omega_0$
obtained in Ref.~\cite{MNRAS} reads $\Omega_0 = 0.7^{+0.8}_{-0.5}$,
which remains consistent with our theoretical result $\Omega_0 = 1$;
see the formula (\ref{om1}).  The most recent Boomerang experiment
\cite{boomerang} indicates that $\ell_{\it peak} = (197 \pm 6)$, which
strongly favours a flat Universe, i.e.\ is a significant confirmation of
the hypothesis that $\Omega_0 \approx 1$.  Thus, it seems that within
the framework of the toy model, the currently observed value of the CMBR
fluctuations remains consistent with the notion of the toy-model
universe in which the value of the cosmological mass density parameter
$\Omega_0$ is equal to one.

\subsubsection{Conclusions}

We can say that the toy model introduces quite significant corrections
to definitions of the most important cosmological parameters,
simultaneously introducing considerable revision of the methodology of
the experimental determination of their values. This may cause, for
instance, that the values of the cosmological density parameters --
obtained by means of different methods from different observational
data -- will be consistent with each other, thereby maintaining the
self-consistency of the model.

In particular, we should draw our attention to the two kinds of
measurements:

{\bf i)} The measured (or calculated with the use of other measured
quantities) value of the cosmological mass density parameter is
considerably smaller than one; it appears that in such a case the
measured (or calculated) quantity could be the cosmological density
parameter $\Omega_0^{\it stand}$ defined as in the standard model; see
section 5.2.1.

{\bf ii)} The measured (or calculated from other measured quantities)
value of the cosmological mass density parameter is close to one; it
turns out that the measured (or calculated) quantity corresponds to the
cosmological density parameter $\Omega_0$ defined as in the toy model,
the value of which, according to the formula (\ref{om1}), is equal to 
one; see section 5.2.3.

\subsection{Early times of the Universe}

The last feature of the toy-model universe that we would like to 
address in this section is its possible lacking of the initial 
singularity, i.e.\ the absence of the singularity at the moment 
$t=0$.(\footnote{Note that some non-singular models of the
Universe are also discussed, for instance, in 
Refs.~\cite{Rosen,Blome,Cooper1,Cooper2,Cooper3,Rebhan}.}) 
First, we observe that as the macrospace ${\cal R}^3$ of the Universe 
is flat -- see the formula (\ref{om1}) -- and then infinite, so it 
existed also for $t \le 0$.
We then assume that $R(t \le 0) = {\it const} \equiv R_s > 0$.  In such
a situation, the spacetime described by the metric (\ref{metryka})
takes the form of the Cartesian (direct) product of the covering
surface ${\cal R}^1 \times {\cal R}^1$ for the anti-de Sitter
two-dimensional universe times the {\it static} flat three-dimensional
macrospace ${\cal R}^3$, and we have
\begin{eqnarray}
ds^2 = \left(1 + |\lambda| a^2\right)\!  c^2 dt^2 - \left( 1 +
|\lambda| a^2\right)^{\! -1} da^2 \nonumber\\ - R^2_s \!\left( d r^2 +
r^2 d \theta^2 + r^2 \sin^2 \theta \, d \varphi^2 \right) \ .
\label{5dim}
\end{eqnarray}
It cannot be ruled out, however, that the primordial toy universe
(before the Big Bang, i.e.\ for $t < 0$) was actually an anti-de Sitter
universe of the form ${\cal S}^1 \times {\cal R}^1 \times {\cal R}^3$,
with closed time-like (or null for radiation) curves, the whole history
of which used to repeat every $T_l \sim 10^{-43}$ seconds.  Such a
universe may be described by the metric
\begin{equation}
ds^2 = c^2 dt^2 - \exp\!\left( i \, 2 \sqrt{|\lambda|} \, c \, t
\right) da^2 
- R^2_s \!\left( d r^2 + r^2 d \theta^2 + r^2 \sin^2 \theta \, d
\varphi^2 \right)
\label{im}
\end{equation}
which we obtain after a purely {\it formal} (complex) coordinate
transformation performed for the metric (\ref{5dim});
in the case of the metric (\ref{im}) the existence 
of the closed time-like (or
null for radiation) curves in the anti-de Sitter universe is
especially easy to notice.  Note that the value of the zero-point 
energy (of the vacuum) in such a spacetime was
exactly the same as in the present toy universe; see section 2.

We will now examine the features of the early Universe in a more
precise manner. Let us first note that the Ricci scalar curvature
${R^\mu}_\mu$ -- present in expression for the action for the Einstein
field equation -- provides the Lagrangian density for the spacetime
geometry represented by the metric tensor \cite{Peebles}.  Let us
suppose the Lagrangian density -- so also the Ricci scalar 
-- to have remained a continuous quantity at the
moment of the Big Bang (i.e.\ for $t=0$); of course, we assume here
that the quantity ${\rm det} (g_{\mu\nu})$ is continuous during
the whole history of the Universe. The Ricci scalar should then
fulfil the following equation,
\begin{equation}
{R^\mu}_\mu \!\left( t \to 0^- \right) = {R^\mu}_\mu \!\left( t \to 0^+
\right) \ .
\label{ccond}
\end{equation}
We assume that the Universe before, after, and at the moment of the Big
Bang was described by the Einstein equation (\ref{1}), with the
stress--energy tensor given by expressions (\ref{tens})--(\ref{ttens2})
and (\ref{t}).  Moreover, we assume that the stress--energy
tensor ${\widehat{T}}_{\mu\nu}$, given by the formulae (\ref{ttens1}) 
and (\ref{ttens2}), was equal to zero before the Big Bang. We will prove
later that this assumption may be valid. From Eq.~(\ref{ccond}) we then
obtain the following relationship between the density $\varrho$ of 
matter, or of the radiation energy 
and its pressures $p$ and $\widetilde{p}$,
\begin{equation}
{\widetilde{p}} + 3 p = c^2 \varrho \ .
\end{equation}
Combining the above equation with Eq.~(\ref{cond}) which is expected to
be fulfilled also in the limit $t \to 0^+$, gives
\begin{eqnarray}
p &=& \frac{c^2}{3} \varrho \\ {\widetilde{p}} &=& 0 \ .
\end{eqnarray}
Thus, in the toy model, the Universe at the moment of the Big Bang was
filled with a perfect relativistic fluid, or radiation. The question
arises as to what was the value of the energy density of the fluid at 
the moment $t=0$, i.e.\ what was the initial condition for the 
density of the radiation energy in the Universe.  To answer this
question, let us imagine that the Universe before the Big Bang was of
the form of the anti-de Sitter spacetime ${\cal S}^1 \times {\cal R}^1
\times {\cal R}^3$. We know that the anti-de Sitter spacetime 
${\cal S}^1 \times {\cal R}^1$ contains ``global'' closed time-like or
null curves; see sections 3 and 4.2 of this paper.  Let us assume that
such a universe was filled with the radiation whose particles had a
period of oscillation
\begin{equation}
T_l = \frac{2 \pi}{c\sqrt{|\lambda|}} \ ,
\label{tl}
\end{equation}
so they retraced their own life histories after each lapse of the
period $T_l$ of the coordinate time $t$.  Thus, the geodesic lines 
of the radiation particles (quanta) were closed null curves, 
so the radiation
was ``frozen'', or ``fixed'' in time $t$ in the anti-de Sitter
spacetime; see Fig.~2.  Note that the distance covered -- during the
coordinate time equal to the period of oscillation $T_l$ -- by the 
radiation particles (quanta) in the ``macroscopic'' space ${\cal R}^3$
was equal to $L = c \, T_l$, which quantity we can call 
a ``wavelength'' of oscillation. The ``frozen'' radiation in the 
macrospace ${\cal R}^3$ might be described, or imagined in such a way 
that each quantum of the radiation covered the distance $L$, equal to 
its (one) wavelength, during the coordinate time $T_l$,
equal to the period of its oscillation; note that, due to the existence
of the closed curves in the investigated anti-de Sitter spacetime, the
time $T_l$ may be regarded as the whole period, or interval of time $t$
which elapsed before the Big Bang; see Fig.~2. It should be added here
that in the two above sentences as well as in the remaining part of
this paper, under the terms ``macroscopic'' space, or macrospace 
${\cal R}^3$ we understand mainly the three-dimensional space given by 
the expression $a \approx 0$.

\bigskip

What happened at the moment of the Big Bang? It seems that the anti-de
Sitter two-dimensional spacetime ${\cal S}^1 \times {\cal R}^1$ should
have undergone a phase transition into the covering surface ${\cal R}^1
\times {\cal R}^1$ of the anti-de Sitter spacetime, described by the 
metric (\ref{3}), so the coordinate time $t$ was released to elapse
monotonically and not periodically as before the Big Bang. This caused
that the radiation started to give a (non-vanishing) contribution to
the stress--energy tensor ${\widetilde{T}}_{\mu\nu}$, in the form
represented by the tensor ${\widehat{T}}_{\mu\nu}$; see the formula 
(\ref{tens}). This, in turn, gave the beginning to the expansion of
the Universe, i.e.\ of the flat three-dimensional space ${\cal R}^3$,
according to Eqs.~(\ref{r1r})--(\ref{r3r}). 
Simultaneously, the radiation, confined before the Big Bang within the 
anti-de Sitter two-dimensional spacetime and forming ``frozen waves'' 
in the macrospace ${\cal R}^3$, was released into the expanding
three-dimensional ``macroscopic'' space.  The Big Bang event occurring
in the toy model could then be described as the following
``spacetime''-- or rather only ``time'' -- transition,
\begin{equation} 
{\cal S}^1 \times {\cal R}^1 \times {\cal R}^3
\stackrel{t=0}{\longrightarrow}
{\cal R}^1 \times {\cal R}^1 \times \!\left( {\cal R}^3
\right)_{\! {\it expand}} \ .
\label{Big}
\end{equation}

What was the energy density of the radiation at the moment of the 
Big Bang? The energy of one quantum of the radiation confined before 
the Big Bang within the anti-de Sitter two-dimensional spacetime is 
easy to calculate,
\begin{equation}
E = \hbar \frac{2\pi}{T_l} = \hbar \sqrt{|\lambda|} \, c \ ;
\label{energy}
\end{equation}
it is, in fact, equal to the Planck energy as defined in section 4.2.
We have to multiply the energy $E$ by a factor of $3$, if we want to
take into account the three orthogonal to each other as well as
independent of -- or ``non-interfering'' with -- one another spatial
modes of the radiation, as the latter filled the three-dimensional
macrospace ${\cal R}^3$.  Furthermore, the so-obtained number should
then be multiplied by a factor of $4$, since the existence of any
trajectory $(t, a, x, y, z)[\gamma]$ of the radiation quantum in the
spacetime described by the metric (\ref{5dim}) or (\ref{im}) implies,
for e.g.\ $y, z = {\it const}$, the existence of the trajectories:\ $(t,
-a, x, y, z)[\gamma]$ and $(t, a, -x, y, z)[\gamma]$ as well as $(t,
-a, -x, y, z)[\gamma]$; see also the formula (\ref{r1}) for comparison.
Note, however, that here we do not take into account the symmetry 
$\gamma \leftrightarrow - \gamma$ of the investigated solution to the 
Einstein equation (\ref{1}), as both kinds
of the trajectories:\ $(\cdot)[\gamma]$ and $(\cdot)[-\gamma] = 
- (\cdot)[\gamma]$, remain the
same in the spacetime of the toy model; it does not, of course, 
exclude the existence of antiparticles -- 
see the last sentence of the discussion
concerning antiparticles in section 3.1 of this paper.
Thus, bearing in mind the above discussion concerning the
closed null curves and the retracing of the life history by the
radiation quanta, one can assume that (for $a=0$) the (twelve) quanta
possessing the energy $12 E$ occupied the region $L \times L \times L$
of the macrospace ${\cal R}^3$. If we then conventionally define the
energy density $u$ as a quotient of the energy concentrated in some
region of the (three-dimensional) space by the volume of this region,
then we will obtain the energy density of the radiation 
at the moment of the Big Bang,
\begin{equation}
u (t \le 0, a=0) = \frac{12 E}{L^3} = \frac{3 \hbar \lambda^2 c}{2
\pi^3} \cong 2.242 \times 10^{112} \;\, {\rm kg} \;\, {\rm s}^{-2}
\;\; {\rm m}^{-1}.
\label{denin}
\end{equation}
As the macrospace ${\cal R}^3$ was/is flat during the whole history of
the toy universe, so consequently one has $\Omega = 1$, then we
obtain
\begin{equation}
H(t \to 0^+) = \left[ \frac{8 \pi G}{3 c^2} u(t \le 0, a=0) \right]^{\!
1/2} \cong 1.181 \times 10^{43} \;\, {\rm s}^{-1}.
\label{hin}
\end{equation}

Note that after the Big Bang (occurring at the moment $t = t_s = 0$) 
the energy density of the radiation should, 
according to Eqs.~(\ref{rr1}), (\ref{rc})
and (\ref{eqeq}), have been decreasing following the formula
\begin{equation}
u(t, a) = \left[ 2 X \!\left( t - t_s \right) +
\left(\frac{u_s}{A}\right)^{-1/2}\right]^{\! -2} 
\end{equation}
where $X \equiv (- \kappa c^2 A/3)^{1/2}$ and $u_s \equiv u(t = t_s,
a = 0)$. It is easy to see -- compare, for instance, the formulae
(\ref{5dim}) and (\ref{2}) as the two solutions to formally the same
Einstein equation with the same stress--energy
tensor -- that the initial condition for the scale factor $R$ may be
e.g.\ expressed as follows,
\begin{equation}
R_s \equiv R( t \le 0 ) = 1 \; {\rm m},
\end{equation}
which allows one to determine the value of the integration constant
resulting from the continuity equation for radiation, 
$u R^4 = u_s R_s^4 = {\it const}$, as well as to obtain the value of
the quantity $\dot{R} (t \to 0^+)$, from Eq.~(\ref{rr1}). We should 
note that if we assume the cosmic scale factor $R$ to be an 
``order parameter'' for the toy-model universe,
then we can recognize the process described by expression 
(\ref{Big}) as a phase transition of the first order, since the 
quantity $R$ remained continuous at the moment of the 
Big Bang, whereas its derivative $\dot{R}$ was not continuous 
at the time $t = 0$, as one has
$\dot{R} (t \to 0^-) = 0 \neq \dot{R} (t \to 0^+)$; of course,
this happened because exactly at the moment of the Big Bang 
the radiation filling the toy universe just started to give 
a (non-vanishing) contribution to 
the stress--energy tensor ${\widetilde{T}}_{\mu\nu}$.

\bigskip

Now we will try to show that the stress--energy tensor
${\widehat{T}}_{\mu\nu}$ was equal to zero before the Big Bang. To this
aim let us assume that the smallest element of the fluid (``frozen''
radiation), which the toy universe before the Big Bang was filled with,
was -- in the macrospace ${\cal R}^3$ -- the cube of the sizes $L
\times L \times L$.(\footnote{One could not 
imagine a smaller element of the fluid than that of the sizes $L \times
L \times L$, since each of the three sizes of the smallest fluid
element should contain at least one full length of the
oscillations of the
radiation quantum, i.e.\ exactly the quantity $L$.}) The cube was filled
with the four -- symmetrical with respect to the four 
hypersurfaces $a = 0$
and $x, y, z = {\it const}$ as well as non-interfering with each other
-- sets of the three independent of and orthogonal to one another
spatial modes of the ``frozen'' radiation, each of the wavelength $L =
c \, T_l$.  The sizes of the (hyper)cube in the directions $t$ and $a$
were equal to $T_l \equiv L/c$ and $L/\pi$ [see the formula (\ref{d_3})],
respectively. We should now recall the general definition of a
stress--energy tensor $T^{\mu\nu}$, which determines the element
$T^{\mu\nu}$ for a particular $\mu$ and $\nu$ as the flux density of
the $\mu$-component of momentum in the direction $\nu$, i.e.\ through
the surface with a constant value of the coordinate $x^{\nu}$,
orthogonally to it.  It is then clear that the flux density of any
component $\mu = 0, \ldots, 4$ of momentum through any of the surfaces 
-- with constant values of the coordinates $x^\nu$ 
(for all $\nu = 0, \ldots, 4$) -- 
of the hyperbox $c \, T_l \times L/\pi \times L
\times L \times L$ was actually equal to zero before 
the Big Bang.\footnote{It is of particular interest to observe that the
above conclusion holds also for the surfaces of the considered
hyperbox, which are given by $a = {\it const}$, or $a = \pm L/(2 \pi)$,
and $t = {\it const} \pm n T_l$, where $n \in {\cal N} \cup \{ 0 \}$.} 
This happened due to the fact that before the Big Bang the geodesic 
lines of the radiation quanta were the closed null curves with 
the closure times (or the
oscillation periods) all equal to $T_l$, so none of the surfaces of the
considered hyperbox was crossed by the radiation quanta during the
period of time $T_l$, and/so the time $T_l$ may actually be regarded as
the whole period, or interval of time $t$ which elapsed before the Big
Bang.  Let us note that the above considerations seem to be true not
only for the element $c \, T_l \times L/\pi \times L \times L \times L$
of the fluid filling the toy universe before the Big Bang, but also
remain valid for any multiplication of this element, of the form $n_1 c
\, T_l \times n_2 L/\pi \times n_3 L \times n_4 L \times n_5 L$, 
where all $n_1, \ldots, n_5$ are natural numbers.

\bigskip
\smallskip

What happened later? After the lapse of a some time after the Big Bang,
the perfect relativistic fluid, or radiation -- expanding
within/together with the three-dimensional ``macroscopic'' space ${\cal
R}^3$ -- started to convert into ``ordinary'' matter, for which the
relative ratio $p/u$ decreases from the value $1/3$ to $0$, and thus,
according to Eq.~(\ref{cond}), the pressure $\widetilde{p}$ along
the extra spatial dimension $a$ decreases from $0$ to $- u/2$.

It is important to observe that, apart from the attempt to answer the
question about the initial conditions at the moment of the Big Bang,
the toy model seems to enable us also to solve the problem of the
large-scale spatial homogeneity and isotropy of the present Universe.
Namely, the toy-model universe existed already {\it before} the Big
Bang and each (separate) region $L \times L \times L$ of the
``macroscopic'' (flat) three-dimensional space ${\cal R}^3$ before the
Big Bang was filled with the radiation of the same ``initial'' energy
$12 E$, where the quantity $E$ is given by the formula (\ref{energy}). 
We may then understand that the energy density of the radiation 
filling the early Universe was highly isotropic already at the moment of
the Big Bang. The possible fluctuations in the density of energy/matter,
which started to grow some time after the Big Bang, could have been
caused by the fact that the energy $U$ at the moment of the Big Bang
could have varied for different quanta of the radiation.

\section{Summary}

In this paper we have investigated the toy model which originates from
general relativity. It provides an extension of the dimensionality of
spacetime, with an additional dimension of space macroscopically
unobservable. The model attempts to give a solution to the problem of
the cosmological constant.

It turns out that the toy model introduces no corrections to most
predictions of the ``standard'' general relativity regarding, among 
others, the so-called ``five tests of general relativity''. However, 
it seems that the toy model could provide an explanation to the
flatness of circular velocity curves of spiral galaxies without 
introducing any dark matter.

The toy model introduces certain changes into cosmology, altering the
definition of the critical density of matter. Consequently, it also
changes -- as compared with the Friedmann standard model -- the 
values assumed by other cosmological parameters. Due to
the introduction of the additional spatial dimension, data concerning
the present value of the Universe's mass density obtained, for
instance, from observations of distant supernovae seem to be consistent
with other measurements, such as those regarding the temperature
fluctuations of the cosmic microwave background radiation. Finally, no
initial singularity is present in the proposed model.

\bigskip
\bigskip
\noindent {\bf Acknowledgments}
\bigskip

\noindent
The author is most grateful to {\it Alexander von Humboldt-Stiftung}
for support and for making possible his research stay in Germany, where
studies on the subject discussed in this paper were started. 
He is also indebted to Professor W.~Kopczy\'nski for his critical 
remarks regarding the very preliminary version of sections 2 and 3 
of this essay.

\newpage

\centerline{\epsfxsize=5.6 true in \epsfbox{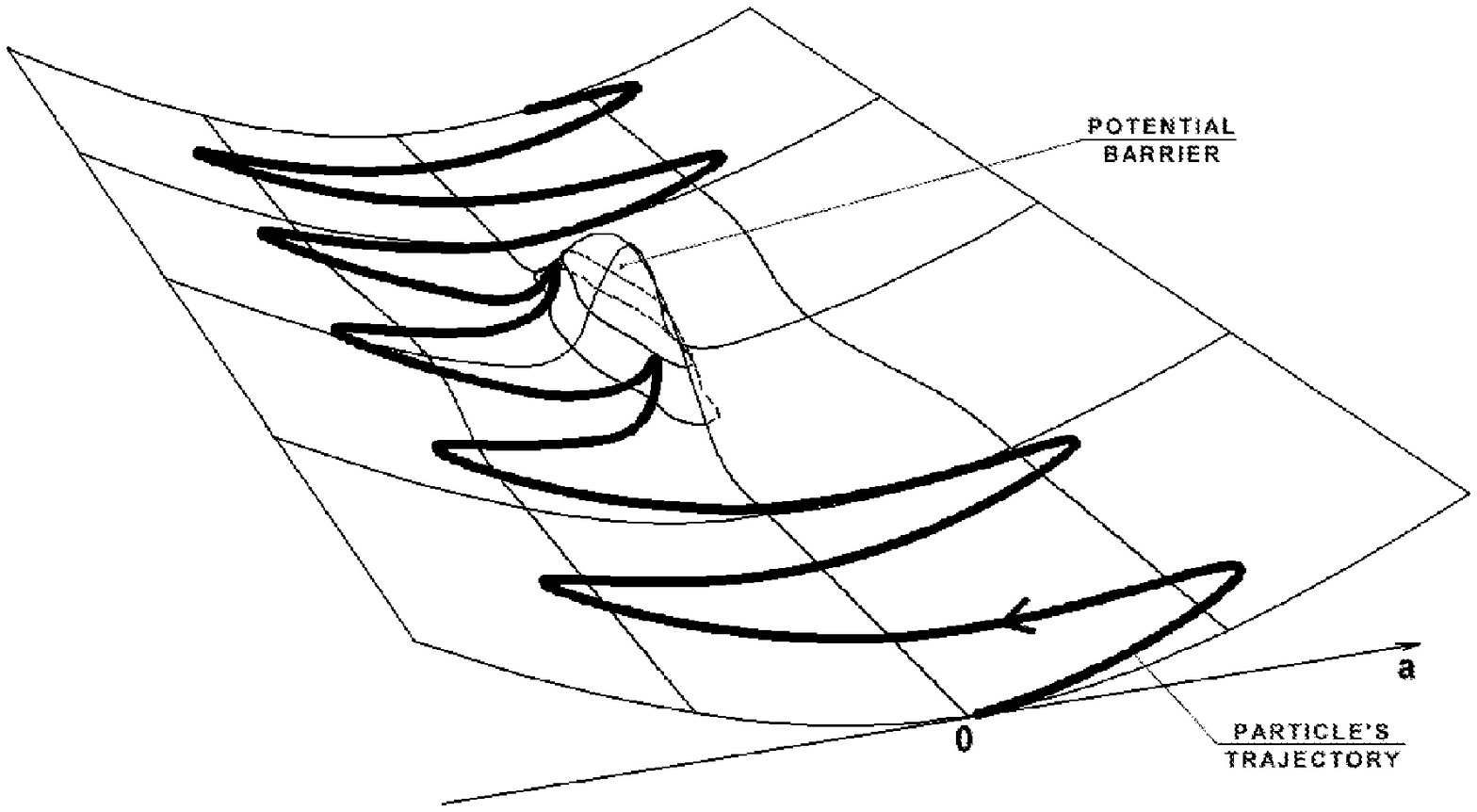}}

\bigskip
\bigskip
\smallskip

\noindent {\small {\bf Figure 1.}\ A particle, 
moving with a uniform rectilinear motion
with respect to the spacetime $a = 0$, simultaneously oscillates in
the additional fourth spatial dimension $a$. The particle encounters an
obstacle -- a potential barrier higher than its energy $E$. However,
because of the oscillations in the fourth spatial dimension, the particle
can surmount the barrier, by simply circumventing it. The nature of
this phenomenon is purely geometric and thus extremely simple. We can
see here an analogy to a tunnelling through the potential barrier, the
phenomenon known from quantum mechanics and still puzzling -- as far as
its nature is concerned.  According to the corpuscular interpretation
of quantum mechanics, at the moment of the tunnelling the particle
temporarily ``vanishes'' to appear subsequently on the opposite side of
the barrier. In the toy model presented in this paper, the particle can
``vanish'' in the extra spatial dimension, which may allow it to
circumvent the potential barrier.

Let us note that the particle can cross the potential barrier (or
rather circumvent it) as well as be reflected.  Naturally, the answer
to the question of whether a particle will cross (circumvent) the
potential barrier or whether it will rebound, depends on a number of
factors such as:\ the values of the particle's energies $U$ and $E$, the
barrier sizes as well as the barrier shape with respect to the extra
spatial dimension $a$. {{\bf{\em Note.}}\ The figure is merely 
demonstrative and does not preserve proportions between various 
quantities characterizing both the particle and the potential barrier.}

\newpage

\centerline{\epsfxsize=3.00 true in \epsfbox{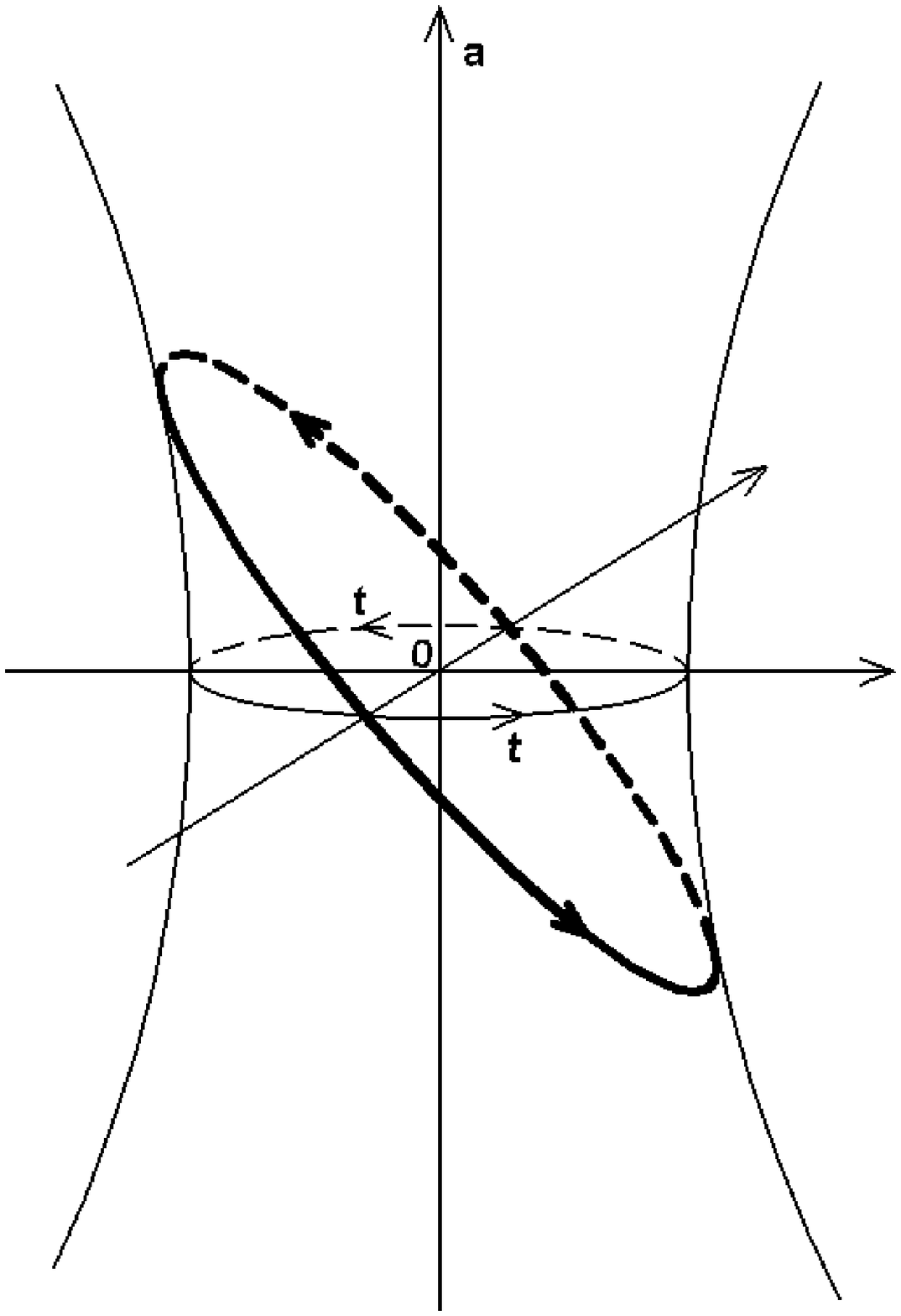}}

\medskip

\noindent {\small {\bf Figure 2.}\ The toy model 
predicts that the Universe before the Big
Bang was of the form of the anti-de Sitter spacetime ${\cal S}^1
\times {\cal R}^1 \times {\cal R}^3$ 
filled with the radiation quanta, each of energy $E$ equal to the
Planck energy $E_{\it Pl}$. The geodesic line(s) of the radiation
particle(s), or quanta (the thick line in the figure) were global
closed null curves -- with the periods of oscillation all equal to the
Planck time $T_l$ -- so the radiation was ``frozen'' in time $t$.  In
the macrospace ${\cal R}^3$, each quantum of the ``frozen'' radiation
covered the distance $L$, equal to its one wavelength, during the time
$T_l$ which may be regarded as the whole period, or interval of time $t$
which elapsed before the Big Bang.  At the moment of the Big Bang, the
phase transition ${\cal S}^1 \to {\cal R}^1$ of the coordinate time $t$
occurred, which gave the beginning to the expansion of the flat
three-dimensional space ${\cal R}^3$ and to the 
release of the radiation into the expanding macrospace ${\cal R}^3$.
{{\bf{\em Note.}}\ For reasons of clarity, 
only the anti-de Sitter two-dimensional
spacetime ${\cal S}^1 \times {\cal R}^1$ is shown, so the figure
does not incorporate the flat space ${\cal R}^3$. Note
also that the figure represents the orthogonal projection of 
the trajectory of a single radiation particle -- moving in the
spacetime ${\cal S}^1 \times {\cal R}^1 \times {\cal R}^3$, 
in a one direction of the macrospace ${\cal R}^3$ -- onto the spacetime
${\cal S}^1 \times {\cal R}^1$.}

\end{document}